\documentclass[prl, twocolumn, floatfix, superscriptaddress, longbibliography, reprint]{revtex4-2}
\usepackage[pdftex,plainpages=false,colorlinks=true,linkcolor=blue, citecolor=blue, urlcolor=blue]{hyperref}
\usepackage{amsfonts}
\usepackage{amsmath}
\usepackage{amssymb}
\usepackage{graphicx}
\usepackage{natbib}
\usepackage{color}
\usepackage{placeins}
\usepackage{physics}
\DeclareMathAlphabet\mathbfcal{OMS}{cmsy}{b}{n} 

\begin{document}
 
\title{Dynamics of superconducting pairs in the two-dimensional Hubbard model}
\author{G. Sordi}
\email[corresponding author: ]{giovanni.sordi@rhul.ac.uk}
\affiliation{Department of Physics, Royal Holloway, University of London, Egham, Surrey, UK, TW20 0EX}
\author{E. M. O'Callaghan}
\affiliation{Department of Physics, Royal Holloway, University of London, Egham, Surrey, UK, TW20 0EX}
\author{C. Walsh}
\affiliation{Department of Physics, Royal Holloway, University of London, Egham, Surrey, UK, TW20 0EX}
\author{M. Charlebois}
\affiliation{D\'epartement de Biochimie, Chimie, Physique et Science Forensique, Institut de Recherche sur l'Hydrog\`ene, Universit\'e du Qu\'ebec \`a Trois-Rivi\`eres, Trois-Rivi\`eres, Qu\'ebec, Canada, G9A 5H7}
\author{P. S\'emon}
\affiliation{D\'epartement de physique, Institut quantique \& RQMP, Universit\'e de Sherbrooke, Sherbrooke, Qu\'ebec, Canada J1K 2R1}
\author{A.-M. S. Tremblay}
\affiliation{D\'epartement de physique, Institut quantique \& RQMP, Universit\'e de Sherbrooke, Sherbrooke, Qu\'ebec, Canada J1K 2R1}
\date{\today}

\begin{abstract} 
The frequency structure of the superconducting correlations in cuprates gives insights on the pairing mechanism. Here we present an exhaustive study of this problem in the two-dimensional Hubbard model with cellular dynamical mean-field theory. To this end, we systematically quantify the dependence on doping $\delta$ and interaction strength $U$ of the superconducting gap, of the frequency scales where $d$-wave pairing occurs, and of their relative contribution to pairing. For all values of $U$ and $\delta$, we find pair-forming processes confined to frequencies set by the superexchange interaction and followed by pair-breaking processes, ruling out both pair-forming and pair-breaking processes on the scale of $U$. This suggests that at high frequencies, the effect of $U$ is eliminated by the $d$-wave paring, and that at small frequencies, $U$ generates the superexchange interaction that leads to low-frequency pair-forming processes providing the net contribution to pairing.
\end{abstract}
 
\maketitle

{\it Introduction}-- 
The mechanism driving the superconducting pairing of electrons in cuprates is still under scrutiny~\cite{Norman2011, Scalapino_RMP2012, keimerRev}. It requires knowing how the electrons forming the Cooper pairs are correlated {\it both} in space and in time. On one hand, the $d$-wave symmetry of the superconducting gap and the short correlation length measured in cuprates suggest that the {\it spatial dependence} of the pair correlations is nonlocal and short-ranged~\cite{keimerRev}. On the other hand, the limited experimental information about the {\it time dependence} of the pair correlations in cuprates~\cite{Heumen:PRB2009, Carbotte_2011, DalConte:Science2012, Cilento_2013, DalConte:NatPhys2015} hinders a thorough characterisation of the pairing mechanism~\cite{Anderson:Science2007, Scalapino_RMP2012}.

While first-principle methods now allow material specific predictions~\cite{Cedric:NatPhys2010, Cedric:EPL2012, Acharya:PRX2018, Bacq:PRX2025, Cui:NatComm2025}, the detailed information of model calculations can provide valuable insights for clarifying the universal features of the pairing mechanism and for guiding experimental progress. 
These calculations are even more pressing in view of current proposals for measuring pair correlations with time-resolved~\cite{Boschini:RMP2024} and coincidence~\cite{Kemper_2025, Devereaux:PRB2023, Stahl:PRB2019, Su:PRB2020} angle-resolved photoemission spectroscopy (ARPES). 
The strong electron-electron interaction in cuprates indicates that the two-dimensional (2D) Hubbard model, where electrons hop in a square lattice with an amplitude $t$ and experience an onsite Coulomb repulsion $U$, is the point of departure for modeling the superconducting pairing~\cite{Anderson:1987, Scalapino_RMP2012}. However, clarifying the origin of pairing in the Hubbard model is still a theoretical challenge, owing to the nonperturbative nature of the strong electronic correlations~\cite{AMJulich}.

Regarding the {\it spatial dependence} of the superconducting correlations, the Hubbard model correctly captures their nonlocal character. Physically, this is because the repulsion $U$ (i) disfavors the occurrence of two electrons on the same site and (ii) dynamically generates the antiferromagnetic superexchange interaction $J=4t^2/U$ which favors antiparallel spins on neighboring sites, and thus their effective attraction~\cite{Morel:PRB1962, Kotliar:PRB1988}. Detailed studies confirm these effects brought about by $U$~\cite{maier, tremblayR, QinAnnuRev2022, AMJulich, Scalapino_RMP2012}. 

The issue of the {\it time dependence}, i.e. of the dynamics, of the superconducting correlations in the Hubbard model has been less explored and is controversial. It is the focus of this work. This is a difficult task, requiring a theory able to treat on an equal footing different and coexisting time scales, from short time scales associated with the interaction $U$ to longer time scales associated with the superexchange $J$. Cluster extensions~\cite{kotliarRMP, maier, tremblayR} of dynamical mean-field theory~\cite{rmp}, which handle both spatial fluctuations (within the cluster) and all temporal fluctuations, provide such a theory. 

The key questions concern the characteristic time scales that lead to pairing and their relative contribution to pairing. These features depend on $U$ and doping. Previous studies~\cite{maierPRL2008, Kyung:2009, civelli1, senechalPRB2013, Gull:PRB2014, reymbautPRB2016, DongPNAS2022, DongNatPhys2022} show that the main contribution to pairing arises from relatively long time (low frequency) scales ascribed to short-range spin singlet correlations generated by the superexchange $J$, with smaller contributions arising from shorter time (higher frequency) scales linked to $U$. However, these conclusions are drawn from studies on a selected range of parameters of the 2D Hubbard model and are based on different methodologies. For example, the pairing dynamics was studied in Ref.~\cite{maierPRL2008} with dynamical cluster approximation (DCA) at $U/t=8, 10, 12$ and $20\%$ doping at finite temperature $T$, in Ref.~\cite{Kyung:2009} with cellular dynamical mean-field theory (CDMFT) at $U/t=8$ and few doping levels at $T=0$, in Refs.~\cite{Gull:PRB2014, DongPNAS2022, DongNatPhys2022} with DCA at $U/t=5.5, 6$ and few doping levels at low $T$, and in Ref.~\cite{reymbautPRB2016} with CDMFT at $U/t=9$ and several doping levels at low $T$. 

Here we revisit the problem of the dynamics of the superconducting pairs in the 2D Hubbard model using state of the art calculations based on CDMFT~\cite{kotliarRMP, maier, tremblayR} at finite temperature. Methodologically, the added value of our work is twofold. First, taking advantage of algorithmic improvements and large investment of computing time, we explore a dataset encompassing a wide range of interaction $U$ and a comprehensive group of doping levels. Second, our findings  are a direct computational result based solely on the Green’s function obtained in CDMFT, with no other assumptions coming e.g. from low-frequency theories. 
Using this approach, we systematically quantify the dependence on doping and $U$ of the characteristic frequency scales where pairing occurs and their relative contribution to pairing. For all values of doping and $U$, we find a rich structure in the dynamics of the correlations of the paired electrons, with pair-forming processes mainly confined on the energy scale set by $J$ and followed by pair-breaking processes. Hence, the net contribution to pairing comes only from the low frequency pair-forming processes.

{\it Model and Method}-- 
We study the 2D Hubbard model on the square lattice, $H = - \sum_{ij \sigma} t_{ij} c_{i\sigma}^{\dagger} c_{j\sigma} + U\sum_{i} n_{i\uparrow} n_{i\downarrow} -\mu \sum_{i\sigma} n_{i\sigma}$. Here, $c_{i\sigma}$ ($c_{i\sigma}^{\dagger}$) is the operator that destroys (creates) an electron with spin $\sigma=\{ \uparrow, \downarrow \}$ at site $i$, $n=c_{i\sigma}^{\dagger}c_{i\sigma}$ is the number operator, $t_{ij}$ is the hopping amplitude between nearest neighbour sites, $U$ is the onsite Coulomb repulsion, and $\mu$ is the chemical potential which changes the occupation $n$ and thus the hole doping $\delta=1-n$. We set $t_{ij}=t=1$ as our energy unit.

We solve this model in the $d_{x^2-y^2}$-wave superconducting state and at finite temperature with CDMFT~\cite{kotliarRMP, maier, tremblayR}. CDMFT maps the Hubbard model onto a cluster impurity model embedded in a self-consistent bath of noninteracting electrons. Here we consider the minimal cluster that describes $d$-wave superconductivity, i.e. a $2\times 2$ plaquette. To solve the cluster impurity model, we use the hybridization expansion continuous-time quantum Monte Carlo method (CT-HYB)~\cite{hauleCTQMC, Werner:2006, millisRMP} with a LazySkip List algorithm~\cite{patrickSkipList} and with Monte Carlo updates of two pairs of creation and destruction operators to ensure ergodicity~\cite{patrickERG}. The CT-HYB method enables us to explore the superconducting state for a wide range of interaction strength $U$ and to reach low temperatures.
As shown by extensive studies, CDMFT on a $2\times 2$ plaquette can describe several properties of the superconducting state that are in qualitative agreement with experiments~\cite{tremblayR, AMJulich, QinAnnuRev2022, Sakai:JPSJ2023} and robust against the cluster size~\cite{QinAnnuRev2022, Liu:CPL2025}. They include the dome-like shape of the superconducting state~\cite{kancharla, hauleDOPING, sshtSC, LorenzoSC, Hebert:2015} and its interplay with the pseudogap~\cite{kancharla, senechalAFSC2005, hauleDOPING, sshtSC, LorenzoSC}, the features of the density of states and of the spin susceptibility~\cite{hauleDOPING, Civelli:PRL2008, Walsh:PRB2023}, and the changes in energy and entropy upon condensation~\cite{carbone2006, LorenzoSC, CaitlinPNAS2021}. 

This work focuses on the dynamics of the superconducting pairs, which is encoded in the Green’s function. In the cluster momentum basis, the Green’s function is 
\begin{align}
G_{\bf K}(\tau) & = \left( \begin{array}{cc}
G_{{\bf K} \uparrow}(\tau) & F_{\bf K}(\tau) \\
F_{\bf K}^{+}(\tau) & -G_{-{\bf K} \downarrow}(-\tau)
\end{array} \right) , 
\label{eq:G}
\end{align}
where, introducing the Matsubara frequencies, $G_{{\bf K}\sigma}(i\omega_n) = -\int_0^\beta d\tau e^{i\omega_n \tau} \langle T_\tau c_{{\bf K}\sigma}(\tau) c_{{\bf K}\sigma}^\dagger(0)\rangle$ is the Nambu diagonal (i.e. normal) Green’s function and $F_{{\bf K}}(i\omega_n) = -\int_0^\beta d\tau e^{i\omega_n \tau} \langle T_\tau c_{{\bf K}\uparrow}(\tau) c_{-{\bf K}\downarrow}(0)\rangle$ is the Nambu off-diagonal (i.e. anomalous) Green’s function. For $d$-wave superconductivity $F_{{\bf K}=(\pi,0)}=-F_{{\bf K}=(0,\pi)}$ is the only nonzero component. The superconducting order parameter is $\Phi=\langle F_{{\bf K}=(\pi,0)}(\tau=0^{+}) \rangle$. 

We perform the analytical continuation from imaginary to real frequencies using the maximum entropy software of Ref.~\cite{DominicMEM}. However, for the anomalous Green’s function, which shows both positive and negative spectral weight, the direct maximum entropy method is inapplicable. Instead, we use the MaxEntAux method of Ref.~\cite{Alexis_PRB2015}, which relies on an auxiliary Green’s function with positive spectral weight, and exploit the symmetry properties of the anomalous Green’s function (see supplemental material~\cite{SupplementalMaterial}).

\begin{figure*}
\centering{
\includegraphics[width=0.94\linewidth]{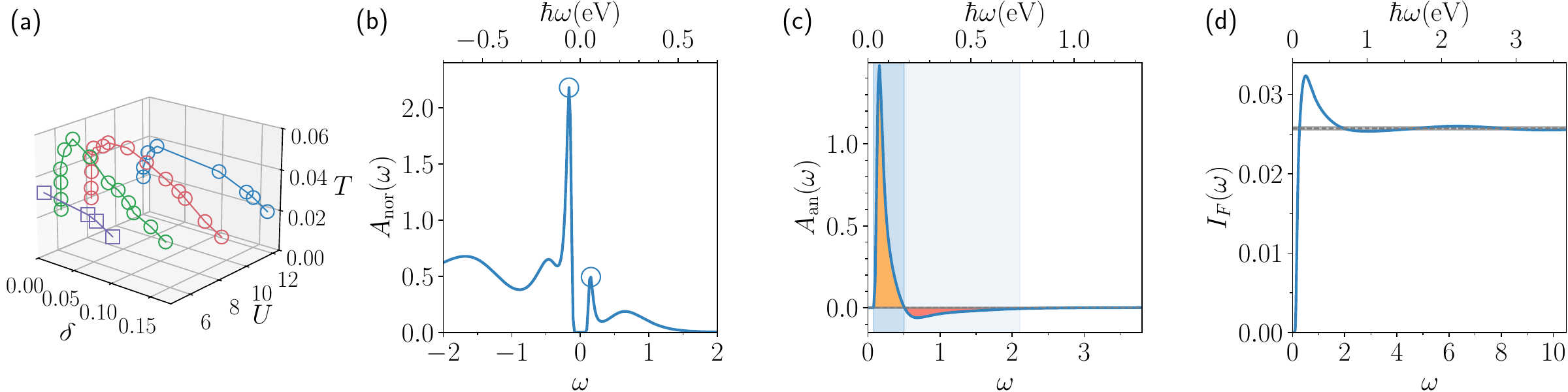}
}
\caption{(a) Superconducting transition temperature $T_c^{\rm CDMFT}$ versus $\delta$ for different values of $U$ (data taken from Ref.~\cite{CaitlinPNAS2021}). (b) Density of states in the superconducting state $A_{\rm nor}(\omega)$. The half distance between the position of the coherence peaks (marked by open circles) gives the superconducting gap $\Delta_{\rm sc}$ and is analysed in Fig.~\ref{fig:SCgap}. (c) Anomalous spectral function $A_{\rm an}(\omega)$. The frequency regions where $A_{\rm an}(\omega)$ is positive (negative) are those that contribute to pairing (depairing): they are shaded with dark (pale) blue vertical bands and are examined in Fig.~\ref{fig:freq-pairing}. The areas where $A_{\rm an}(\omega)$ is positive or negative are colored and analysed in Fig.~\ref{fig:areas}. (d) Cumulative spectral weight of the order parameter $I_F(\omega)$. Horizontal line denotes the superconducting order parameter $|\Phi|$ computed independently. Data in panels (b), (c) and (d) are for $U=12$, $T=1/50$ and $\delta \approx 0.02$. Frequencies are converted into physical units with $t=350$~meV. 
}
\label{fig:intro}
\end{figure*}

{\it Strategy}-- 
The strategy of this work is to systematically examine the dependence on doping and $U$ of three key features of the superconducting state: (i) the superconducting gap, (ii) the characteristic frequency scales that contribute to superconducting pairing and (iii) their relative contribution to pairing.

To do that, we first take advantage of previous work that mapped out the superconducting state of the 2D Hubbard model with CDMFT on a $2\times 2$ plaquette in the $U-\delta-T$ space~\cite{hauleDOPING, sshtSC, LorenzoSC, CaitlinPNAS2021}. This is condensed in Fig.~\ref{fig:intro}(a), where we report the data of Ref.~\cite{CaitlinPNAS2021} showing the superconducting transition temperature $T_c^{\rm CDMFT}$ (defined as the temperature below which $|\Phi|$ is nonzero~\footnote{This neglects Kosterlitz-Thouless physics}) as a function of $\delta$ for different values of $U$ straddling the normal state metal to Mott insulator transition at $U_{\rm MIT} \approx 5.95$~\cite{CaitlinSb}. Below $U_{\rm MIT}$ (squares), $T_c^{\rm CDMFT}(\delta)|_U$ reaches its highest value at $\delta=0$ and monotonically decreases with increasing doping. Above $U_{\rm MIT}$ (circles), $T_c^{\rm CDMFT}(\delta)|_U$ acquires an asymmetric dome-like shape versus $\delta$, reaching its highest value at a finite doping which is dependent on $U$. As a function of $U$, $T_c^{\rm CDMFT}$ is optimised just above $U_{\rm MIT}$ (green circles). 

Next, we fix the temperature at $T=1/50$, since it is below the optimum $T_c^{\rm CDMFT}$ for each value of $U$, and we explore the model for $U \in [ 5.2, 16]$ and several doping levels. For each value of $U$ and $\delta$ we calculate the density of states in the superconducting state $A_{\rm nor}(\omega) = -\frac{1}{\pi}{\rm Im} G_{\rm {\bf R}=(0,0)}(\omega)$, the anomalous spectral function $A_{\rm an}(\omega)=-\frac{1}{\pi}{\rm Im} F_{\rm {\bf K}=(\pi,0)}(\omega)$, and the cumulative spectral weight of the order parameter~\cite{Kyung:2009, reymbautPRB2016} $I_F(\omega)=\int_{-\omega}^{\omega} \frac{d\omega'}{2\pi} A_{\rm an} (\omega') f({-\omega'})$. Fig.~\ref{fig:intro}(b),(c),(d) show $A_{\rm nor}(\omega)$, $A_{\rm an}(\omega)$ and $I_F(\omega)$ for $U=12$ and $\delta \approx 0.02$, as a sample of our calculations (see supplemental Figs.~S3-S7 for spectra at other model parameters).

From $A_{\rm nor}(\omega)$, we extract a key feature of the superconducting state, i.e. the superconducting gap $2\Delta_{\rm sc}$. It is the minimum energy to break a Cooper pair. We estimate $\Delta_{\rm sc}$ as the half distance between the position of the coherence peaks in $A_{\rm nor}(\omega)$ [circles in Fig.~\ref{fig:intro}(b)].

From $A_{\rm an}(\omega)$, which for $d$-wave superconductivity is real and odd in frequency~\cite{Alexis_PRB2015}, we extract key features of the dynamics of the superconducting pairs, namely the frequency intervals where pairing occurs and their relative contribution to pairing. To do that, first we extract the frequency intervals where $A_{\rm an}(\omega)$ is positive [shaded dark blue region in Fig.~\ref{fig:intro}(c)]. These are the frequencies that contribute to pairing processes (i.e. pair-forming). Similarly, the frequencies over which $A_{\rm an}(\omega)$ is negative contribute to depairing processes (i.e. pair-breaking) [shaded pale blue region in Fig.~\ref{fig:intro}(c)]. Second, we calculate the corresponding areas between $A_{\rm an}(\omega)$ and the frequency axis [colored areas in Fig.~\ref{fig:intro}(c)] to identify the relative contribution to pairing for each frequency range, as explained in the following paragraph. 

To show that positive and negative anomalous spectral weight determines the frequency range where pairing and depairing arise, we can turn to the behavior of the cumulative spectral weight of the order parameter $I_{F}(\omega)$ [Fig.~\ref{fig:intro}(d)]. In the limit $\omega \rightarrow \infty$, $I_F(\omega)$ converges to the superconducting order parameter $|\Phi|$ (horizontal grey line) [see also supplemental Fig.~S2]. At low temperatures, $I_F(\omega)$ is approximately the integral of $A_{\rm an}(\omega)$ over the positive frequencies, i.e. $I_F(\omega) \approx \int_{0}^{\omega} \frac{d\omega'}{2\pi} A_{\rm an} (\omega')$. Hence, positive anomalous spectral weight  $A_{\rm an}(\omega)$ enhances $|\Phi|$ and thus is pair-forming, whereas negative weight depletes $|\Phi|$ and thus is pair-breaking. Hence, we can extract the pair-forming (pair-breaking) frequencies from the frequency ranges where $I_F(\omega)$ is increasing (decreasing). Similarly, we can extract the contribution to pairing (depairing) from the difference between a maximum (minimum) value of $I_F(\omega)$ and its preceding minimum (maximum) value.

This physical interpretation of $A_{\rm an}(\omega)$ and $I_F(\omega)$ has been tested in the BCS and Eliashberg cases in Ref.~\cite{Kyung:2009}. 
\begin{figure}
\centering{
\includegraphics[width=0.94\linewidth]{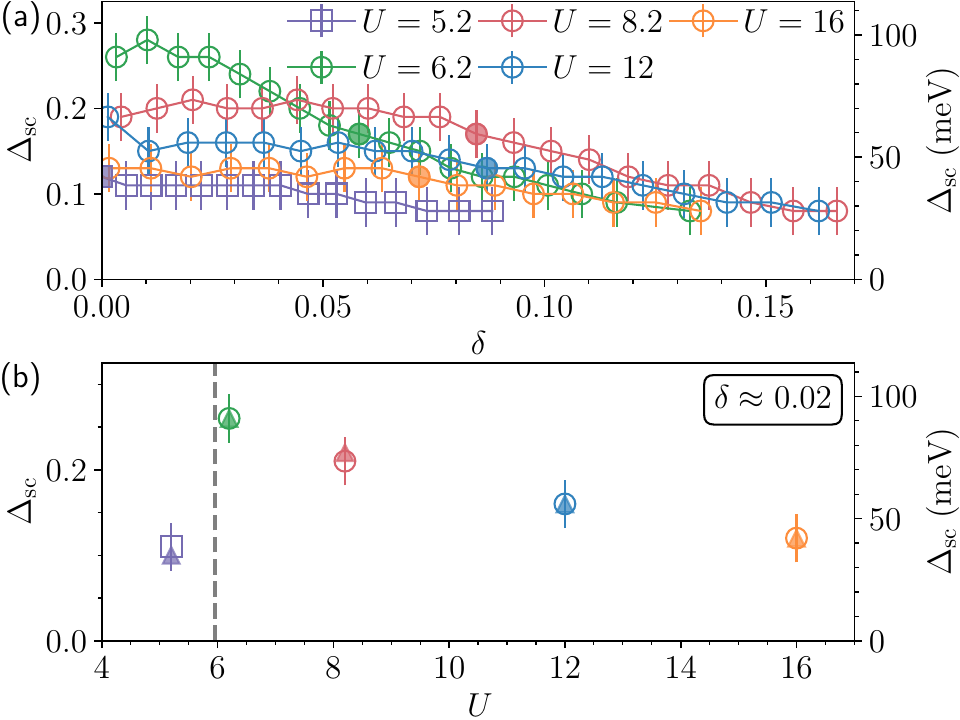}
}
\caption{(a): Superconducting gap $\Delta_{\rm sc}$ versus $\delta$ for different values of $U$, as determined by the half of the frequency difference of the coherence peaks in $A_{\rm nor}(\omega)$, see e.g. Fig. \ref{fig:intro}(b). Error bars indicate the uncertainty in determining the peaks. For each value of $U$, the filled symbol denotes the doping at which the superconducting order parameter $|\Phi(\delta)|$ has a maximum. (b) $\Delta_{\rm sc}$ at fixed doping $\delta \approx 0.02$ versus $U$. The dashed vertical line marks the critical threshold $U_{\rm MIT}\approx 5.95$~\cite{CaitlinSb} for opening the Mott gap at half-filling. Filled up triangles denote the position of the low-frequency peak in $A_{\rm an}(\omega)$, which tracks $\Delta_{\rm sc}$. Data are converted into physical units with $t=350$~meV.
}
\label{fig:SCgap}
\end{figure}

{\it Superconducting gap}-- 
Figure~\ref{fig:SCgap}(a) shows the superconducting gap $\Delta_{\rm sc}$ as a function of doping for different values of $U$. $\Delta_{\rm sc}$ is defined as half of the energy distance between the coherence peaks in $A_{\rm nor}(\omega)$. Physically, it reflects the pairing strength. Starting from high doping, as the doping decreases, $\Delta_{\rm sc}$ increases. For $U>U_{\rm MIT}$, and on decreasing doping further, $\Delta_{\rm sc}$ flattens. Remarkably, for $U>U_{\rm MIT}$, the doping dependence of $\Delta_{\rm sc}$  contrasts with that of the superconducting order parameter $|\Phi|$, which first increases and then decreases as a function of doping [open squares in Fig.~\ref{fig:areas}]. The value of doping for which $|\Phi|$ is maximum, $\delta_{\Phi^{\rm max}}$, is indicated by a filled symbol on each curve. Given that this doping is found~\cite{LorenzoSC} to be larger than the optimal doping (i.e. the doping that maximises $T_c^{\rm CDMFT}(\delta)$), the doping dependence of $\Delta_{\rm sc}$ also contrasts with that of $T_c^{\rm CDMFT}$. Hence, in the underdoped region for $U>U_{\rm MIT}$, $\Delta_{\rm sc}$ flattens while both $T_c^{\rm CDMFT}$ and $|\Phi|$ drop on approaching the Mott insulator at $\delta=0$. This non-BCS behavior is compatible with previous calculations~\cite{Paramekanti:2004, kancharla, Civelli:PRL2008, hauleDOPING, Gull:2013, reymbautPRB2016, Walsh:PRB2023} and with ARPES experiments in hole-doped cuprates~\cite{Sobota:RMP2021}. 

Figure~\ref{fig:SCgap}(b) shows $\Delta_{\rm sc}$ versus $U$ at the fixed low doping $\delta \approx 0.02$. Remarkably, $\Delta_{\rm sc}(U)$ shows a non-monotonic behavior, peaking around the underlying normal state Mott transition at $U_{\rm MIT}$ (dashed vertical line). In contrast, $|\Phi|$ at this doping level decreases monotonically with increasing $U$. 

\begin{figure*}
\centering{
\includegraphics[width=0.94\linewidth]{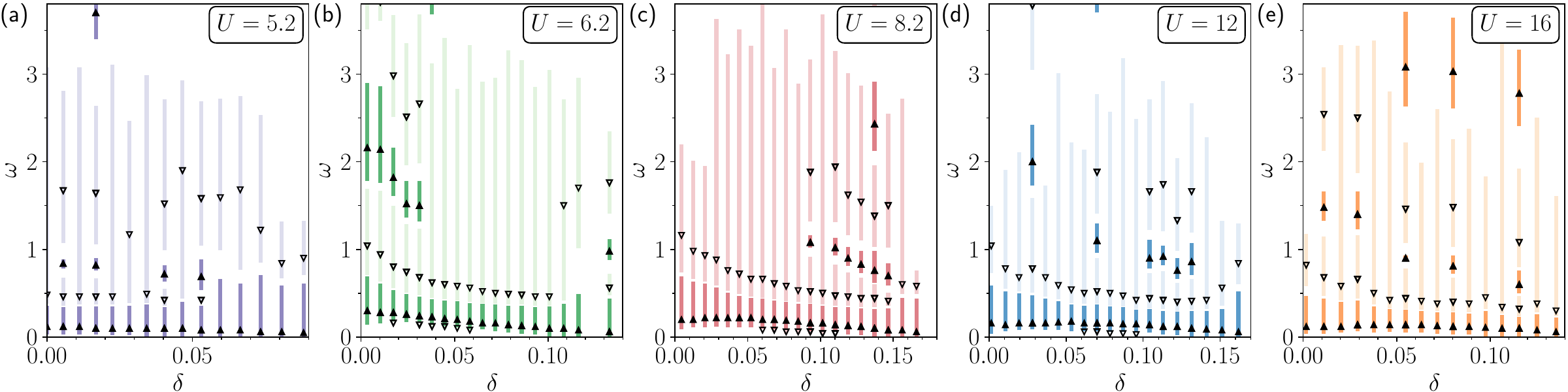}
}
\caption{Vertical dark (pale) bars indicate the frequency regions where $A_{\rm an}(\omega)$ is positive (negative), for several values of doping. Up (down) triangles denote the maxima (minima) in each frequency region. Blank space between these bars indicates the regions where $A_{\rm an}(\omega)$ is negligible (here $|A_{\rm an}(\omega)| <0.005$). Each panel shows data for a given interaction strength $U$.}
\label{fig:freq-pairing}
\end{figure*}

{\it Frequency scales where pairing arises}-- 
Next, we analyse the frequency intervals where pairing occurs. These are given by the regions where $A_{\rm an}(\omega)$ is positive. They are shown in Fig.~\ref{fig:freq-pairing} with vertical and dark colored bars, for several values of $\delta$ and $U$. 
Vertical and pale colored bars indicate the regions where $A_{\rm an}(\omega)$ is negative, i.e. where depairing occurs. Up (down) triangles mark the maxima (minima) in each frequency region. Blank space between these bars indicates the regions where $A_{\rm an}(\omega)$ is negligible (here, $|A_{\rm an}(\omega)| <0.005$). This small treshold has been introduced to filter out the noise in the spectra (see supplemental material).  

Although $A_{\rm an}(\omega)$ for a given value of $U$ and $\delta$ may show a complicated behavior due to the limitations of the analytical continuation, a few trends emerge. As a function of frequency, there is one main interval yielding pair-forming processes alternating with one main interval yielding pair-breaking processes, followed by higher frequency processes that are indistinguishable from noise. 

First, at low frequencies $A_{\rm an}(\omega)$ has a gap (blank space in Fig.~\ref{fig:freq-pairing}) between $\omega=0$ and $\omega \approx 0.1$ (see e.g. Fig.~\ref{fig:intro}(c) and Fig.~\ref{fig:lowfreq-pairing} in Appendix for a low-frequency zoom of Fig.~\ref{fig:freq-pairing}). The behavior of this gap versus $U$ and $\delta$ approximately follows that of $\Delta_{\rm sc}$, i.e. saturates at small doping levels and decreases with increasing doping and with increasing $U$ above $U_{\rm MIT}$. 

Above this gap, $A_{\rm an}(\omega)$ is positive on a narrow frequency range, which for all values of $U$ and $\delta$ lies between $\omega \approx 0$ and $\omega \approx 0.6$ (see e.g. dark shaded blue region in Fig.~\ref{fig:intro}(c)). (i) This frequency scale is of the order of $J/2$, suggesting that the long-lived pair-forming processes occurring in this frequency interval are associated with short-range spin fluctuations, as pointed out earlier~\cite{maierPRL2008, Kyung:2009, Gull:PRB2014}. (ii) This low-frequency region leading to pairing occurs for all doping levels and for all $U$, suggesting that it is unrelated to the underlying normal-state strongly correlated pseudogap, which only develops for $U>U_{\rm MIT}$ and small dopings~\cite{sht, sht2, ssht}. (iii) In this low-frequency region, $A_{\rm an}(\omega)$ shows a prominent peak, whose position tracks $\Delta_{\rm sc}$, both as a function of $U$ (filled up triangles vs open symbols in Fig.~\ref{fig:SCgap}(b)) and $\delta$ (Fig.~\ref{fig:lowfreq-pairing}). As noted in Ref.~\cite{reymbautPRB2016}, this suggests that the optimum spectral weight for pairing in $A_{\rm an}(\omega)$ takes place on a frequency scale close to $\Delta_{\rm sc}$.

Above this low-frequency region leading to pairing, $A_{\rm an}(\omega)$ becomes negative over a broad frequency range, which for all values of $U$ and $\delta$ lies between $\omega \approx 0.6$ and $\omega \approx 3.0$ (see e.g. pale shaded blue region in Fig.~\ref{fig:intro}(c)). This region leading to depairing is not restricted to low doping or large values of $U$. 

For further higher frequencies, $A_{\rm an}(\omega)$ remains overall small and indistinguishable from noise, for all doping levels and for all $U$ (see e.g. blank region in Fig.~\ref{fig:intro}(c) for $\omega \gtrsim 2.1$). Although we cannot rule out that pair-forming and pair-breaking processes may also occur at higher frequency scales, and in particular on the scale of order $U$, our results show that their contribution to pairing is negligible, as we shall discuss in the next section.

\begin{figure*}
\centering{
\includegraphics[width=0.94\linewidth]{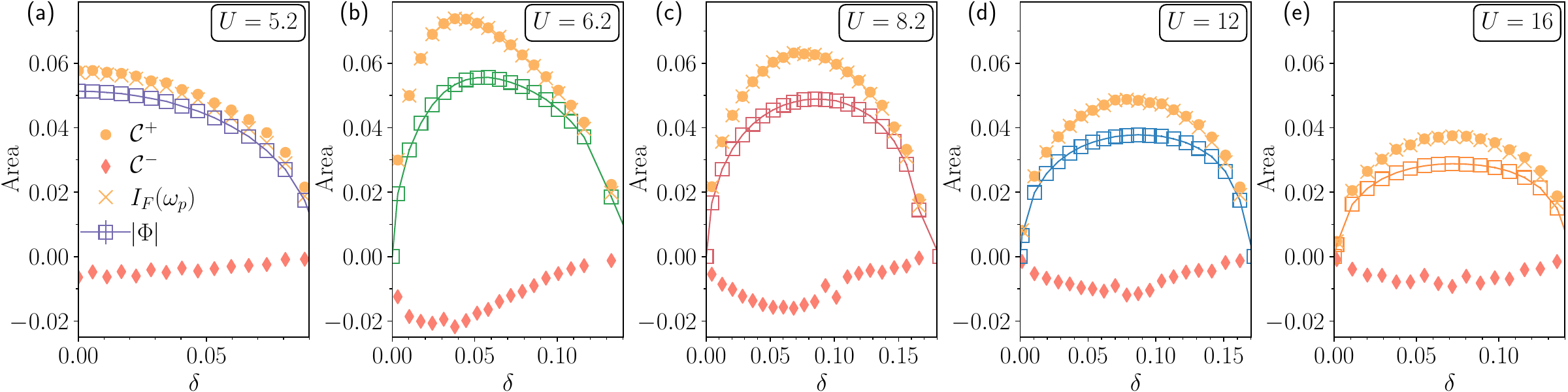}
}
\caption{The largest area, divided by $2\pi$, under the positive regions and above the negative regions of $A_{\rm an}(\omega)$ versus doping $\delta$, for different values of $U$: $\mathcal{C}^{+}$ (positive area, yellow filled circles) and $\mathcal{C}^{-}$ (negative area, red filled diamonds). They are calculated as follows. With the composite trapezoidal method, we find the largest area under (above) $A_{\rm an}(\omega)$ and above (under) the small threshold of $\pm 0.005$, to filter out noise. Yellow crosses indicate the value of the low-frequency peak of the cumulative spectral weight of the order parameter, $I_F(\omega_p)$. They follow $\mathcal{C}^{+}$. Open squares denote the superconducting order parameter $|\Phi|$.}
\label{fig:areas}
\end{figure*}

{\it Contribution to pairing}-- 
Next, we study the contribution to pairing or to depairing for each frequency interval where pairing or depairing occurs. This can be estimated by the area under the positive regions or above the negative regions of $A_{\rm an}(\omega)$ (see e.g. Fig.~\ref{fig:intro}(c)). 
Figure~\ref{fig:areas} shows the largest area, divided by $2\pi$, under the positive regions and above the negative regions of $A_{\rm an}(\omega)$ ($\mathcal{C}^{+}$, yellow filled circles for the largest positive area and $\mathcal{C}^{-}$, red filled diamonds for the largest negative area) versus $\delta$ for different values of $U$. This largest positive (negative) area occurs for the main frequency interval where pairing (depairing) occurs that we have identified in Fig.~\ref{fig:freq-pairing}, i.e. $\mathcal{C}^{+}$ arises from the interval $0 \lesssim \omega \lesssim 0.6$ and $\mathcal{C}^{-}$ arises from the interval $0.6 \lesssim \omega \lesssim 3.0$. We verified that the positive and negative areas at other frequency scales are much smaller or negligible (see supplemental material). This implies the key result that the contribution to pairing (or depairing) of higher frequencies is negligible, for all values of $\delta$ and $U$. Specifically, our results rule out an increased importance of higher frequency pair-forming processes with increasing $U$, a possibility left opened in previous work~\cite{maierPRL2008, Kyung:2009}.  

For $U>U_{\rm MIT}$, $\mathcal{C}^{+}$ has a dome-like shape versus $\delta$ and overall decreases with increasing $U$.  Hence, a key finding is that on approaching the Mott insulator at $\delta=0$, the drop of $|\Phi|$ and $T_c^{\rm CDMFT}$ is associated with the reduction of the area of the low frequency peak of $A_{\rm an}(\omega)$~\cite{Kyung:2009, reymbautPRB2016, Gull:PRB2014}. The behavior of $\mathcal{C}^{-}$ is similar to that of $\mathcal{C}^{+}$, although $\mathcal{C}^{-}$ is negative and is smaller in magnitude than $\mathcal{C}^{+}$. 

Having analysed the contribution to pairing and to depairing for each frequency interval, we examine their {\it net} contribution to pairing. Since we found that there is just one main frequency interval where pairing occurs, the answer is clear: the net contribution to pairing comes only from the low-frequency pair-forming processes.  

Specifically, the net contribution to pairing for a given frequency interval is the signed area of $A_{\rm an}(\omega)$ in that frequency interval. Since we found that only $\mathcal{C}^{+}$ and $\mathcal{C}^{-}$ are sizeable, $\mathcal{C}^{+} +\mathcal{C}^{-}$ is approximately the order parameter $|\Phi|$. Physically, this means that only the low frequencies build up the order parameter. From Fig.~\ref{fig:areas}, $\mathcal{C}^{+}(\delta)$ is greater than $|\Phi(\delta)|$ [open squares], for all values of $U$ and $\delta$. It means $\mathcal{C}^{+}$ is partly canceled out, so that only a fraction of $\mathcal{C}^{+}$ contributes to $|\Phi|$ and thus to pairing. Physically, this means that the low frequency pair-forming processes outweigh the pair-breaking ones, so that only a fraction of the low-frequency pair-forming processes provides a net contribution to pairing. Put another way, the net contribution to pairing comes only from the low-frequency pair-forming processes. 

The partial cancellation of the anomalous spectral weight can immediately be deduced from the behavior of the cumulative spectral weight of the order parameter $I_F(\omega)$ [see e.g. Fig.~\ref{fig:intro}(d)]: at low frequency, $I_F(\omega)$ {\it overshoots} $|\Phi|$. If pairing had a net contribution from higher frequencies, then $|\Phi|$ would be larger than the value of the low frequency peak, $I_F(\omega_p)$. Similarly, Fig.~\ref{fig:areas} shows that $I_F(\omega_p)$ [yellow crosses], which basically coincides with $\mathcal{C}^{+}$, is larger than $|\Phi|$ for all values of $U$ and $\delta$.

{\it Summary}-- 
We studied the dynamics of the superconducting pairs in the 2D Hubbard model for a wide range of interaction $U$ and doping. The superconducting gap $\Delta_{\rm sc}$ extracted from $A_{\rm nor}(\omega)$ does not scale with the superconducting order parameter, but instead saturates on approaching the Mott insulator. 
$ A_{\rm an}(\omega)$ has a rich structure as a function of frequency. (i) For all values of $U$ and $\delta$, the pair-forming processes, described by the positive weight of $A_{\rm an}(\omega)$, are mainly confined on a single frequency interval set by the superexchange. (ii) They are followed by pair-breaking processes, described by the negative weight of $A_{\rm an}(\omega)$, which are in turn followed by further higher frequency processes that are indistinguishable from noise.
The contribution to pairing and depairing can be extracted from the areas under the positive and negative regions of $A_{\rm an}(\omega)$. (i) For all values of $U$ and $\delta$, the contribution to pairing of higher frequencies is negligible, ruling out pair-forming and pair-breaking processes on the scale of $U$. This suggests that $d$-wave pairing, that creates a node in the pair wave function when both electrons are on the same site, suffices to eliminate the effect of $U$ at high frequency, which would be pair-breaking in the $s$-wave channel. 
(ii) Since the low frequency pair-forming processes outweigh the pair-breaking ones, the net contribution to pairing comes only from the low-frequency frequency pair-forming processes, which occurs on the scale of the superexchange $J$. This reflects that at small frequencies, $U$ dynamically generates the interaction $J$, which plays a role analogous to the Debye frequency in the case of phonons, where pairing processes occur below that frequency and pair-breaking processes occur above that frequencies. 
Our results provide predictions for experiments that plan to directly measure pair correlations with time-resolved~\cite{Boschini:RMP2024}  and coincidence~\cite{Kemper_2025, Devereaux:PRB2023, Stahl:PRB2019, Su:PRB2020} ARPES.

\begin{acknowledgments}
{\it Acknowledgments}-- 
This work has been partially supported by the Canada First Research Excellence Fund. Simulations were performed on computers provided by the Canada Foundation for Innovation, Calcul Qu\'ebec, and Digital Research Alliance of Canada. 
\end{acknowledgments}

\bibliography{pairing}

\begin{thebibliography}{64}%
\makeatletter
\providecommand \@ifxundefined [1]{%
 \@ifx{#1\undefined}
}%
\providecommand \@ifnum [1]{%
 \ifnum #1\expandafter \@firstoftwo
 \else \expandafter \@secondoftwo
 \fi
}%
\providecommand \@ifx [1]{%
 \ifx #1\expandafter \@firstoftwo
 \else \expandafter \@secondoftwo
 \fi
}%
\providecommand \natexlab [1]{#1}%
\providecommand \enquote  [1]{``#1''}%
\providecommand \bibnamefont  [1]{#1}%
\providecommand \bibfnamefont [1]{#1}%
\providecommand \citenamefont [1]{#1}%
\providecommand \href@noop [0]{\@secondoftwo}%
\providecommand \href [0]{\begingroup \@sanitize@url \@href}%
\providecommand \@href[1]{\@@startlink{#1}\@@href}%
\providecommand \@@href[1]{\endgroup#1\@@endlink}%
\providecommand \@sanitize@url [0]{\catcode `\\12\catcode `\$12\catcode
  `\&12\catcode `\#12\catcode `\^12\catcode `\_12\catcode `\%12\relax}%
\providecommand \@@startlink[1]{}%
\providecommand \@@endlink[0]{}%
\providecommand \url  [0]{\begingroup\@sanitize@url \@url }%
\providecommand \@url [1]{\endgroup\@href {#1}{\urlprefix }}%
\providecommand \urlprefix  [0]{URL }%
\providecommand \Eprint [0]{\href }%
\providecommand \doibase [0]{https://doi.org/}%
\providecommand \selectlanguage [0]{\@gobble}%
\providecommand \bibinfo  [0]{\@secondoftwo}%
\providecommand \bibfield  [0]{\@secondoftwo}%
\providecommand \translation [1]{[#1]}%
\providecommand \BibitemOpen [0]{}%
\providecommand \bibitemStop [0]{}%
\providecommand \bibitemNoStop [0]{.\EOS\space}%
\providecommand \EOS [0]{\spacefactor3000\relax}%
\providecommand \BibitemShut  [1]{\csname bibitem#1\endcsname}%
\let\auto@bib@innerbib\@empty
\bibitem [{\citenamefont {Norman}(2011)}]{Norman2011}%
  \BibitemOpen
  \bibfield  {author} {\bibinfo {author} {\bibfnamefont {M.~R.}\ \bibnamefont
  {Norman}},\ }\bibfield  {title} {\bibinfo {title} {The challenge of
  unconventional superconductivity},\ }\href
  {https://doi.org/10.1126/science.1200181} {\bibfield  {journal} {\bibinfo
  {journal} {Science}\ }\textbf {\bibinfo {volume} {332}},\ \bibinfo {pages}
  {196} (\bibinfo {year} {2011})}\BibitemShut {NoStop}%
\bibitem [{\citenamefont {Scalapino}(2012)}]{Scalapino_RMP2012}%
  \BibitemOpen
  \bibfield  {author} {\bibinfo {author} {\bibfnamefont {D.~J.}\ \bibnamefont
  {Scalapino}},\ }\bibfield  {title} {\bibinfo {title} {A common thread: The
  pairing interaction for unconventional superconductors},\ }\href
  {https://doi.org/10.1103/RevModPhys.84.1383} {\bibfield  {journal} {\bibinfo
  {journal} {Rev. Mod. Phys.}\ }\textbf {\bibinfo {volume} {84}},\ \bibinfo
  {pages} {1383} (\bibinfo {year} {2012})}\BibitemShut {NoStop}%
\bibitem [{\citenamefont {Keimer}\ \emph {et~al.}(2015)\citenamefont {Keimer},
  \citenamefont {Kivelson}, \citenamefont {Norman}, \citenamefont {Uchida},\
  and\ \citenamefont {Zaanen}}]{keimerRev}%
  \BibitemOpen
  \bibfield  {author} {\bibinfo {author} {\bibfnamefont {B.}~\bibnamefont
  {Keimer}}, \bibinfo {author} {\bibfnamefont {S.~A.}\ \bibnamefont
  {Kivelson}}, \bibinfo {author} {\bibfnamefont {M.~R.}\ \bibnamefont
  {Norman}}, \bibinfo {author} {\bibfnamefont {S.}~\bibnamefont {Uchida}},\
  and\ \bibinfo {author} {\bibfnamefont {J.}~\bibnamefont {Zaanen}},\
  }\bibfield  {title} {\bibinfo {title} {From quantum matter to
  high-temperature superconductivity in copper oxides},\ }\href
  {https://doi.org/10.1038/nature14165} {\bibfield  {journal} {\bibinfo
  {journal} {Nature}\ }\textbf {\bibinfo {volume} {518}},\ \bibinfo {pages}
  {179} (\bibinfo {year} {2015})}\BibitemShut {NoStop}%
\bibitem [{\citenamefont {van Heumen}\ \emph {et~al.}(2009)\citenamefont {van
  Heumen}, \citenamefont {Muhlethaler}, \citenamefont {Kuzmenko}, \citenamefont
  {Eisaki}, \citenamefont {Meevasana}, \citenamefont {Greven},\ and\
  \citenamefont {van~der Marel}}]{Heumen:PRB2009}%
  \BibitemOpen
  \bibfield  {author} {\bibinfo {author} {\bibfnamefont {E.}~\bibnamefont {van
  Heumen}}, \bibinfo {author} {\bibfnamefont {E.}~\bibnamefont {Muhlethaler}},
  \bibinfo {author} {\bibfnamefont {A.~B.}\ \bibnamefont {Kuzmenko}}, \bibinfo
  {author} {\bibfnamefont {H.}~\bibnamefont {Eisaki}}, \bibinfo {author}
  {\bibfnamefont {W.}~\bibnamefont {Meevasana}}, \bibinfo {author}
  {\bibfnamefont {M.}~\bibnamefont {Greven}},\ and\ \bibinfo {author}
  {\bibfnamefont {D.}~\bibnamefont {van~der Marel}},\ }\bibfield  {title}
  {\bibinfo {title} {{Optical determination of the relation between the
  electron-boson coupling function and the critical temperature in
  high-${T}_{c}$ cuprates}},\ }\href
  {https://doi.org/10.1103/PhysRevB.79.184512} {\bibfield  {journal} {\bibinfo
  {journal} {Phys. Rev. B}\ }\textbf {\bibinfo {volume} {79}},\ \bibinfo
  {pages} {184512} (\bibinfo {year} {2009})}\BibitemShut {NoStop}%
\bibitem [{\citenamefont {Carbotte}\ \emph {et~al.}(2011)\citenamefont
  {Carbotte}, \citenamefont {Timusk},\ and\ \citenamefont
  {Hwang}}]{Carbotte_2011}%
  \BibitemOpen
  \bibfield  {author} {\bibinfo {author} {\bibfnamefont {J.~P.}\ \bibnamefont
  {Carbotte}}, \bibinfo {author} {\bibfnamefont {T.}~\bibnamefont {Timusk}},\
  and\ \bibinfo {author} {\bibfnamefont {J.}~\bibnamefont {Hwang}},\ }\bibfield
   {title} {\bibinfo {title} {{Bosons in high-temperature superconductors: an
  experimental survey}},\ }\href
  {https://doi.org/10.1088/0034-4885/74/6/066501} {\bibfield  {journal}
  {\bibinfo  {journal} {Reports on Progress in Physics}\ }\textbf {\bibinfo
  {volume} {74}},\ \bibinfo {pages} {066501} (\bibinfo {year}
  {2011})}\BibitemShut {NoStop}%
\bibitem [{\citenamefont {Conte}\ \emph {et~al.}(2012)\citenamefont {Conte},
  \citenamefont {Giannetti}, \citenamefont {Coslovich}, \citenamefont
  {Cilento}, \citenamefont {Bossini}, \citenamefont {Abebaw}, \citenamefont
  {Banfi}, \citenamefont {Ferrini}, \citenamefont {Eisaki}, \citenamefont
  {Greven}, \citenamefont {Damascelli}, \citenamefont {van~der Marel},\ and\
  \citenamefont {Parmigiani}}]{DalConte:Science2012}%
  \BibitemOpen
  \bibfield  {author} {\bibinfo {author} {\bibfnamefont {S.~D.}\ \bibnamefont
  {Conte}}, \bibinfo {author} {\bibfnamefont {C.}~\bibnamefont {Giannetti}},
  \bibinfo {author} {\bibfnamefont {G.}~\bibnamefont {Coslovich}}, \bibinfo
  {author} {\bibfnamefont {F.}~\bibnamefont {Cilento}}, \bibinfo {author}
  {\bibfnamefont {D.}~\bibnamefont {Bossini}}, \bibinfo {author} {\bibfnamefont
  {T.}~\bibnamefont {Abebaw}}, \bibinfo {author} {\bibfnamefont
  {F.}~\bibnamefont {Banfi}}, \bibinfo {author} {\bibfnamefont
  {G.}~\bibnamefont {Ferrini}}, \bibinfo {author} {\bibfnamefont
  {H.}~\bibnamefont {Eisaki}}, \bibinfo {author} {\bibfnamefont
  {M.}~\bibnamefont {Greven}}, \bibinfo {author} {\bibfnamefont
  {A.}~\bibnamefont {Damascelli}}, \bibinfo {author} {\bibfnamefont
  {D.}~\bibnamefont {van~der Marel}},\ and\ \bibinfo {author} {\bibfnamefont
  {F.}~\bibnamefont {Parmigiani}},\ }\bibfield  {title} {\bibinfo {title}
  {{Disentangling the Electronic and Phononic Glue in a High-$T_c$
  Superconductor}},\ }\href {https://doi.org/10.1126/science.1216765}
  {\bibfield  {journal} {\bibinfo  {journal} {Science}\ }\textbf {\bibinfo
  {volume} {335}},\ \bibinfo {pages} {1600} (\bibinfo {year}
  {2012})}\BibitemShut {NoStop}%
\bibitem [{\citenamefont {Cilento}\ \emph {et~al.}(2013)\citenamefont
  {Cilento}, \citenamefont {Conte}, \citenamefont {Coslovich}, \citenamefont
  {Banfi}, \citenamefont {Ferrini}, \citenamefont {Eisaki}, \citenamefont
  {Greven}, \citenamefont {Damascelli}, \citenamefont {Marel}, \citenamefont
  {Parmigiani},\ and\ \citenamefont {Giannetti}}]{Cilento_2013}%
  \BibitemOpen
  \bibfield  {author} {\bibinfo {author} {\bibfnamefont {F.}~\bibnamefont
  {Cilento}}, \bibinfo {author} {\bibfnamefont {S.~D.}\ \bibnamefont {Conte}},
  \bibinfo {author} {\bibfnamefont {G.}~\bibnamefont {Coslovich}}, \bibinfo
  {author} {\bibfnamefont {F.}~\bibnamefont {Banfi}}, \bibinfo {author}
  {\bibfnamefont {G.}~\bibnamefont {Ferrini}}, \bibinfo {author} {\bibfnamefont
  {H.}~\bibnamefont {Eisaki}}, \bibinfo {author} {\bibfnamefont
  {M.}~\bibnamefont {Greven}}, \bibinfo {author} {\bibfnamefont
  {A.}~\bibnamefont {Damascelli}}, \bibinfo {author} {\bibfnamefont {D.~v.~d.}\
  \bibnamefont {Marel}}, \bibinfo {author} {\bibfnamefont {F.}~\bibnamefont
  {Parmigiani}},\ and\ \bibinfo {author} {\bibfnamefont {C.}~\bibnamefont
  {Giannetti}},\ }\bibfield  {title} {\bibinfo {title} {{In search for the
  pairing glue in cuprates by non-equilibrium optical spectroscopy}},\ }\href
  {https://doi.org/10.1088/1742-6596/449/1/012003} {\bibfield  {journal}
  {\bibinfo  {journal} {Journal of Physics: Conference Series}\ }\textbf
  {\bibinfo {volume} {449}},\ \bibinfo {pages} {012003} (\bibinfo {year}
  {2013})}\BibitemShut {NoStop}%
\bibitem [{\citenamefont {Dal~Conte}\ \emph {et~al.}(2015)\citenamefont
  {Dal~Conte}, \citenamefont {Vidmar}, \citenamefont {Golež}, \citenamefont
  {Mierzejewski}, \citenamefont {Soavi}, \citenamefont {Peli}, \citenamefont
  {Banfi}, \citenamefont {Ferrini}, \citenamefont {Comin}, \citenamefont
  {Ludbrook}, \citenamefont {Chauviere}, \citenamefont {Zhigadlo},
  \citenamefont {Eisaki}, \citenamefont {Greven}, \citenamefont {Lupi},
  \citenamefont {Damascelli}, \citenamefont {Brida}, \citenamefont {Capone},
  \citenamefont {Bonča}, \citenamefont {Cerullo},\ and\ \citenamefont
  {Giannetti}}]{DalConte:NatPhys2015}%
  \BibitemOpen
  \bibfield  {author} {\bibinfo {author} {\bibfnamefont {S.}~\bibnamefont
  {Dal~Conte}}, \bibinfo {author} {\bibfnamefont {L.}~\bibnamefont {Vidmar}},
  \bibinfo {author} {\bibfnamefont {D.}~\bibnamefont {Golež}}, \bibinfo
  {author} {\bibfnamefont {M.}~\bibnamefont {Mierzejewski}}, \bibinfo {author}
  {\bibfnamefont {G.}~\bibnamefont {Soavi}}, \bibinfo {author} {\bibfnamefont
  {S.}~\bibnamefont {Peli}}, \bibinfo {author} {\bibfnamefont {F.}~\bibnamefont
  {Banfi}}, \bibinfo {author} {\bibfnamefont {G.}~\bibnamefont {Ferrini}},
  \bibinfo {author} {\bibfnamefont {R.}~\bibnamefont {Comin}}, \bibinfo
  {author} {\bibfnamefont {B.~M.}\ \bibnamefont {Ludbrook}}, \bibinfo {author}
  {\bibfnamefont {L.}~\bibnamefont {Chauviere}}, \bibinfo {author}
  {\bibfnamefont {N.~D.}\ \bibnamefont {Zhigadlo}}, \bibinfo {author}
  {\bibfnamefont {H.}~\bibnamefont {Eisaki}}, \bibinfo {author} {\bibfnamefont
  {M.}~\bibnamefont {Greven}}, \bibinfo {author} {\bibfnamefont
  {S.}~\bibnamefont {Lupi}}, \bibinfo {author} {\bibfnamefont {A.}~\bibnamefont
  {Damascelli}}, \bibinfo {author} {\bibfnamefont {D.}~\bibnamefont {Brida}},
  \bibinfo {author} {\bibfnamefont {M.}~\bibnamefont {Capone}}, \bibinfo
  {author} {\bibfnamefont {J.}~\bibnamefont {Bonča}}, \bibinfo {author}
  {\bibfnamefont {G.}~\bibnamefont {Cerullo}},\ and\ \bibinfo {author}
  {\bibfnamefont {C.}~\bibnamefont {Giannetti}},\ }\bibfield  {title} {\bibinfo
  {title} {{Snapshots of the retarded interaction of charge carriers with
  ultrafast fluctuations in cuprates}},\ }\href
  {https://doi.org/10.1038/nphys3265} {\bibfield  {journal} {\bibinfo
  {journal} {Nature Physics}\ }\textbf {\bibinfo {volume} {11}},\ \bibinfo
  {pages} {421–426} (\bibinfo {year} {2015})}\BibitemShut {NoStop}%
\bibitem [{\citenamefont {Anderson}(2007)}]{Anderson:Science2007}%
  \BibitemOpen
  \bibfield  {author} {\bibinfo {author} {\bibfnamefont {P.~W.}\ \bibnamefont
  {Anderson}},\ }\bibfield  {title} {\bibinfo {title} {Is there glue in cuprate
  superconductors?},\ }\href {https://doi.org/10.1126/science.1140970}
  {\bibfield  {journal} {\bibinfo  {journal} {Science}\ }\textbf {\bibinfo
  {volume} {316}},\ \bibinfo {pages} {1705} (\bibinfo {year}
  {2007})}\BibitemShut {NoStop}%
\bibitem [{\citenamefont {{Weber}}\ \emph {et~al.}(2010)\citenamefont
  {{Weber}}, \citenamefont {{Haule}},\ and\ \citenamefont
  {{Kotliar}}}]{Cedric:NatPhys2010}%
  \BibitemOpen
  \bibfield  {author} {\bibinfo {author} {\bibfnamefont {C.}~\bibnamefont
  {{Weber}}}, \bibinfo {author} {\bibfnamefont {K.}~\bibnamefont {{Haule}}},\
  and\ \bibinfo {author} {\bibfnamefont {G.}~\bibnamefont {{Kotliar}}},\
  }\bibfield  {title} {\bibinfo {title} {{Strength of correlations in electron-
  and hole-doped cuprates}},\ }\href {https://doi.org/10.1038/nphys1706}
  {\bibfield  {journal} {\bibinfo  {journal} {Nature Physics}\ }\textbf
  {\bibinfo {volume} {6}},\ \bibinfo {pages} {574} (\bibinfo {year}
  {2010})}\BibitemShut {NoStop}%
\bibitem [{\citenamefont {Weber}\ \emph {et~al.}(2012)\citenamefont {Weber},
  \citenamefont {Yee}, \citenamefont {Haule},\ and\ \citenamefont
  {Kotliar}}]{Cedric:EPL2012}%
  \BibitemOpen
  \bibfield  {author} {\bibinfo {author} {\bibfnamefont {C.}~\bibnamefont
  {Weber}}, \bibinfo {author} {\bibfnamefont {C.}~\bibnamefont {Yee}}, \bibinfo
  {author} {\bibfnamefont {K.}~\bibnamefont {Haule}},\ and\ \bibinfo {author}
  {\bibfnamefont {G.}~\bibnamefont {Kotliar}},\ }\bibfield  {title} {\bibinfo
  {title} {{Scaling of the transition temperature of hole-doped cuprate
  superconductors with the charge-transfer energy}},\ }\href
  {https://doi.org/10.1209/0295-5075/100/37001} {\bibfield  {journal} {\bibinfo
   {journal} {Europhysics Letters}\ }\textbf {\bibinfo {volume} {100}},\
  \bibinfo {pages} {37001} (\bibinfo {year} {2012})}\BibitemShut {NoStop}%
\bibitem [{\citenamefont {Acharya}\ \emph {et~al.}(2018)\citenamefont
  {Acharya}, \citenamefont {Weber}, \citenamefont {Plekhanov}, \citenamefont
  {Pashov}, \citenamefont {Taraphder},\ and\ \citenamefont
  {Van~Schilfgaarde}}]{Acharya:PRX2018}%
  \BibitemOpen
  \bibfield  {author} {\bibinfo {author} {\bibfnamefont {S.}~\bibnamefont
  {Acharya}}, \bibinfo {author} {\bibfnamefont {C.}~\bibnamefont {Weber}},
  \bibinfo {author} {\bibfnamefont {E.}~\bibnamefont {Plekhanov}}, \bibinfo
  {author} {\bibfnamefont {D.}~\bibnamefont {Pashov}}, \bibinfo {author}
  {\bibfnamefont {A.}~\bibnamefont {Taraphder}},\ and\ \bibinfo {author}
  {\bibfnamefont {M.}~\bibnamefont {Van~Schilfgaarde}},\ }\bibfield  {title}
  {\bibinfo {title} {{Metal-Insulator Transition in Copper Oxides Induced by
  Apex Displacements}},\ }\href {https://doi.org/10.1103/PhysRevX.8.021038}
  {\bibfield  {journal} {\bibinfo  {journal} {Phys. Rev. X}\ }\textbf {\bibinfo
  {volume} {8}},\ \bibinfo {pages} {021038} (\bibinfo {year}
  {2018})}\BibitemShut {NoStop}%
\bibitem [{\citenamefont {Bacq-Labreuil}\ \emph {et~al.}(2025)\citenamefont
  {Bacq-Labreuil}, \citenamefont {Lacasse}, \citenamefont {Tremblay},
  \citenamefont {S\'en\'echal},\ and\ \citenamefont {Haule}}]{Bacq:PRX2025}%
  \BibitemOpen
  \bibfield  {author} {\bibinfo {author} {\bibfnamefont {B.}~\bibnamefont
  {Bacq-Labreuil}}, \bibinfo {author} {\bibfnamefont {B.}~\bibnamefont
  {Lacasse}}, \bibinfo {author} {\bibfnamefont {A.-M.~S.}\ \bibnamefont
  {Tremblay}}, \bibinfo {author} {\bibfnamefont {D.}~\bibnamefont
  {S\'en\'echal}},\ and\ \bibinfo {author} {\bibfnamefont {K.}~\bibnamefont
  {Haule}},\ }\bibfield  {title} {\bibinfo {title} {{Toward an Ab Initio Theory
  of High-Temperature Superconductors: A Study of Multilayer Cuprates}},\
  }\href {https://doi.org/10.1103/PhysRevX.15.021071} {\bibfield  {journal}
  {\bibinfo  {journal} {Phys. Rev. X}\ }\textbf {\bibinfo {volume} {15}},\
  \bibinfo {pages} {021071} (\bibinfo {year} {2025})}\BibitemShut {NoStop}%
\bibitem [{\citenamefont {{Cui}}\ \emph {et~al.}(2025)\citenamefont {{Cui}},
  \citenamefont {{Yang}}, \citenamefont {{T{\"o}lle}}, \citenamefont {{Ye}},
  \citenamefont {{Yuan}}, \citenamefont {{Zhai}}, \citenamefont {{Park}},
  \citenamefont {{Kim}}, \citenamefont {{Zhang}}, \citenamefont {{Lin}},
  \citenamefont {{Berkelbach}},\ and\ \citenamefont
  {{Chan}}}]{Cui:NatComm2025}%
  \BibitemOpen
  \bibfield  {author} {\bibinfo {author} {\bibfnamefont {Z.-H.}\ \bibnamefont
  {{Cui}}}, \bibinfo {author} {\bibfnamefont {J.}~\bibnamefont {{Yang}}},
  \bibinfo {author} {\bibfnamefont {J.}~\bibnamefont {{T{\"o}lle}}}, \bibinfo
  {author} {\bibfnamefont {H.-Z.}\ \bibnamefont {{Ye}}}, \bibinfo {author}
  {\bibfnamefont {S.}~\bibnamefont {{Yuan}}}, \bibinfo {author} {\bibfnamefont
  {H.}~\bibnamefont {{Zhai}}}, \bibinfo {author} {\bibfnamefont
  {G.}~\bibnamefont {{Park}}}, \bibinfo {author} {\bibfnamefont
  {R.}~\bibnamefont {{Kim}}}, \bibinfo {author} {\bibfnamefont
  {X.}~\bibnamefont {{Zhang}}}, \bibinfo {author} {\bibfnamefont
  {L.}~\bibnamefont {{Lin}}}, \bibinfo {author} {\bibfnamefont {T.~C.}\
  \bibnamefont {{Berkelbach}}},\ and\ \bibinfo {author} {\bibfnamefont
  {G.~K.-L.}\ \bibnamefont {{Chan}}},\ }\bibfield  {title} {\bibinfo {title}
  {{Ab initio quantum many-body description of superconducting trends in the
  cuprates}},\ }\href {https://doi.org/10.1038/s41467-025-56883-x} {\bibfield
  {journal} {\bibinfo  {journal} {Nature Communications}\ }\textbf {\bibinfo
  {volume} {16}},\ \bibinfo {eid} {1845} (\bibinfo {year} {2025})}\BibitemShut
  {NoStop}%
\bibitem [{\citenamefont {Boschini}\ \emph {et~al.}(2024)\citenamefont
  {Boschini}, \citenamefont {Zonno},\ and\ \citenamefont
  {Damascelli}}]{Boschini:RMP2024}%
  \BibitemOpen
  \bibfield  {author} {\bibinfo {author} {\bibfnamefont {F.}~\bibnamefont
  {Boschini}}, \bibinfo {author} {\bibfnamefont {M.}~\bibnamefont {Zonno}},\
  and\ \bibinfo {author} {\bibfnamefont {A.}~\bibnamefont {Damascelli}},\
  }\bibfield  {title} {\bibinfo {title} {Time-resolved arpes studies of quantum
  materials},\ }\href {https://doi.org/10.1103/RevModPhys.96.015003} {\bibfield
   {journal} {\bibinfo  {journal} {Rev. Mod. Phys.}\ }\textbf {\bibinfo
  {volume} {96}},\ \bibinfo {pages} {015003} (\bibinfo {year}
  {2024})}\BibitemShut {NoStop}%
\bibitem [{\citenamefont {Kemper}\ \emph {et~al.}(2025)\citenamefont {Kemper},
  \citenamefont {Goto}, \citenamefont {Labib}, \citenamefont {Gauthier},
  \citenamefont {da~Silva~Neto},\ and\ \citenamefont {Boschini}}]{Kemper_2025}%
  \BibitemOpen
  \bibfield  {author} {\bibinfo {author} {\bibfnamefont {A.~F.}\ \bibnamefont
  {Kemper}}, \bibinfo {author} {\bibfnamefont {F.}~\bibnamefont {Goto}},
  \bibinfo {author} {\bibfnamefont {H.~A.}\ \bibnamefont {Labib}}, \bibinfo
  {author} {\bibfnamefont {N.}~\bibnamefont {Gauthier}}, \bibinfo {author}
  {\bibfnamefont {E.~H.}\ \bibnamefont {da~Silva~Neto}},\ and\ \bibinfo
  {author} {\bibfnamefont {F.}~\bibnamefont {Boschini}},\ }\href
  {https://arxiv.org/abs/2505.01504} {\bibinfo {title} {{Observing two-electron
  interactions with correlation-ARPES}}} (\bibinfo {year} {2025}),\ \Eprint
  {https://arxiv.org/abs/2505.01504} {arXiv:2505.01504 [cond-mat.str-el]}
  \BibitemShut {NoStop}%
\bibitem [{\citenamefont {Devereaux}\ \emph {et~al.}(2023)\citenamefont
  {Devereaux}, \citenamefont {Claassen}, \citenamefont {Huang}, \citenamefont
  {Zaletel}, \citenamefont {Moore}, \citenamefont {Morr}, \citenamefont
  {Mahmood}, \citenamefont {Abbamonte},\ and\ \citenamefont
  {Shen}}]{Devereaux:PRB2023}%
  \BibitemOpen
  \bibfield  {author} {\bibinfo {author} {\bibfnamefont {T.~P.}\ \bibnamefont
  {Devereaux}}, \bibinfo {author} {\bibfnamefont {M.}~\bibnamefont {Claassen}},
  \bibinfo {author} {\bibfnamefont {X.-X.}\ \bibnamefont {Huang}}, \bibinfo
  {author} {\bibfnamefont {M.}~\bibnamefont {Zaletel}}, \bibinfo {author}
  {\bibfnamefont {J.~E.}\ \bibnamefont {Moore}}, \bibinfo {author}
  {\bibfnamefont {D.}~\bibnamefont {Morr}}, \bibinfo {author} {\bibfnamefont
  {F.}~\bibnamefont {Mahmood}}, \bibinfo {author} {\bibfnamefont
  {P.}~\bibnamefont {Abbamonte}},\ and\ \bibinfo {author} {\bibfnamefont
  {Z.-X.}\ \bibnamefont {Shen}},\ }\bibfield  {title} {\bibinfo {title}
  {{Angle-resolved pair photoemission theory for correlated electrons}},\
  }\href {https://doi.org/10.1103/PhysRevB.108.165134} {\bibfield  {journal}
  {\bibinfo  {journal} {Phys. Rev. B}\ }\textbf {\bibinfo {volume} {108}},\
  \bibinfo {pages} {165134} (\bibinfo {year} {2023})}\BibitemShut {NoStop}%
\bibitem [{\citenamefont {Stahl}\ and\ \citenamefont
  {Eckstein}(2019)}]{Stahl:PRB2019}%
  \BibitemOpen
  \bibfield  {author} {\bibinfo {author} {\bibfnamefont {C.}~\bibnamefont
  {Stahl}}\ and\ \bibinfo {author} {\bibfnamefont {M.}~\bibnamefont
  {Eckstein}},\ }\bibfield  {title} {\bibinfo {title} {{Noise correlations in
  time- and angle-resolved photoemission spectroscopy}},\ }\href
  {https://doi.org/10.1103/PhysRevB.99.241111} {\bibfield  {journal} {\bibinfo
  {journal} {Phys. Rev. B}\ }\textbf {\bibinfo {volume} {99}},\ \bibinfo
  {pages} {241111} (\bibinfo {year} {2019})}\BibitemShut {NoStop}%
\bibitem [{\citenamefont {Su}\ and\ \citenamefont {Zhang}(2020)}]{Su:PRB2020}%
  \BibitemOpen
  \bibfield  {author} {\bibinfo {author} {\bibfnamefont {Y.}~\bibnamefont
  {Su}}\ and\ \bibinfo {author} {\bibfnamefont {C.}~\bibnamefont {Zhang}},\
  }\bibfield  {title} {\bibinfo {title} {{Coincidence angle-resolved
  photoemission spectroscopy: Proposal for detection of two-particle
  correlations}},\ }\href {https://doi.org/10.1103/PhysRevB.101.205110}
  {\bibfield  {journal} {\bibinfo  {journal} {Phys. Rev. B}\ }\textbf {\bibinfo
  {volume} {101}},\ \bibinfo {pages} {205110} (\bibinfo {year}
  {2020})}\BibitemShut {NoStop}%
\bibitem [{\citenamefont {Anderson}(1987)}]{Anderson:1987}%
  \BibitemOpen
  \bibfield  {author} {\bibinfo {author} {\bibfnamefont {P.~W.}\ \bibnamefont
  {Anderson}},\ }\bibfield  {title} {\bibinfo {title} {{The resonating valence
  bond state in La$_2$CuO$_4$ and superconductivity}},\ }\href
  {https://doi.org/10.1126/science.235.4793.1196} {\bibfield  {journal}
  {\bibinfo  {journal} {Science}\ }\textbf {\bibinfo {volume} {235}},\ \bibinfo
  {pages} {1196} (\bibinfo {year} {1987})}\BibitemShut {NoStop}%
\bibitem [{\citenamefont {Tremblay}(2013)}]{AMJulich}%
  \BibitemOpen
  \bibfield  {author} {\bibinfo {author} {\bibfnamefont {A.-M.~S.}\
  \bibnamefont {Tremblay}},\ }\bibfield  {title} {\bibinfo {title} {Strongly
  correlated superconductivity},\ }in\ \href
  {http://juser.fz-juelich.de/record/137827/files/FZJ-2013-04137.pdf?version=1}
  {\emph {\bibinfo {booktitle} {Emergent Phenomena in Correlated Matter
  Modeling and Simulation}}},\ Vol.~\bibinfo {volume} {3},\ \bibinfo {editor}
  {edited by\ \bibinfo {editor} {\bibfnamefont {E.}~\bibnamefont {Pavarini}},
  \bibinfo {editor} {\bibfnamefont {E.}~\bibnamefont {Koch}},\ and\ \bibinfo
  {editor} {\bibfnamefont {U.}~\bibnamefont {Schollw\"ock}}}\ (\bibinfo
  {publisher} {Verlag des Forschungszentrum},\ \bibinfo {address} {J\"ulich},\
  \bibinfo {year} {2013})\ Chap.~\bibinfo {chapter} {10}\BibitemShut {NoStop}%
\bibitem [{\citenamefont {Morel}\ and\ \citenamefont
  {Anderson}(1962)}]{Morel:PRB1962}%
  \BibitemOpen
  \bibfield  {author} {\bibinfo {author} {\bibfnamefont {P.}~\bibnamefont
  {Morel}}\ and\ \bibinfo {author} {\bibfnamefont {P.~W.}\ \bibnamefont
  {Anderson}},\ }\bibfield  {title} {\bibinfo {title} {{Calculation of the
  Superconducting State Parameters with Retarded Electron-Phonon
  Interaction}},\ }\href {https://doi.org/10.1103/PhysRev.125.1263} {\bibfield
  {journal} {\bibinfo  {journal} {Phys. Rev.}\ }\textbf {\bibinfo {volume}
  {125}},\ \bibinfo {pages} {1263} (\bibinfo {year} {1962})}\BibitemShut
  {NoStop}%
\bibitem [{\citenamefont {Kotliar}\ and\ \citenamefont
  {Liu}(1988)}]{Kotliar:PRB1988}%
  \BibitemOpen
  \bibfield  {author} {\bibinfo {author} {\bibfnamefont {G.}~\bibnamefont
  {Kotliar}}\ and\ \bibinfo {author} {\bibfnamefont {J.}~\bibnamefont {Liu}},\
  }\bibfield  {title} {\bibinfo {title} {{Superexchange mechanism and d-wave
  superconductivity}},\ }\href {https://doi.org/10.1103/PhysRevB.38.5142}
  {\bibfield  {journal} {\bibinfo  {journal} {Phys. Rev. B}\ }\textbf {\bibinfo
  {volume} {38}},\ \bibinfo {pages} {5142} (\bibinfo {year}
  {1988})}\BibitemShut {NoStop}%
\bibitem [{\citenamefont {Maier}\ \emph {et~al.}(2005)\citenamefont {Maier},
  \citenamefont {Jarrell}, \citenamefont {Pruschke},\ and\ \citenamefont
  {Hettler}}]{maier}%
  \BibitemOpen
  \bibfield  {author} {\bibinfo {author} {\bibfnamefont {T.}~\bibnamefont
  {Maier}}, \bibinfo {author} {\bibfnamefont {M.}~\bibnamefont {Jarrell}},
  \bibinfo {author} {\bibfnamefont {T.}~\bibnamefont {Pruschke}},\ and\
  \bibinfo {author} {\bibfnamefont {M.~H.}\ \bibnamefont {Hettler}},\
  }\bibfield  {title} {\bibinfo {title} {Quantum cluster theories},\ }\href
  {https://doi.org/10.1103/RevModPhys.77.1027} {\bibfield  {journal} {\bibinfo
  {journal} {Rev. Mod. Phys.}\ }\textbf {\bibinfo {volume} {77}},\ \bibinfo
  {pages} {1027} (\bibinfo {year} {2005})}\BibitemShut {NoStop}%
\bibitem [{\citenamefont {Tremblay}\ \emph {et~al.}(2006)\citenamefont
  {Tremblay}, \citenamefont {Kyung},\ and\ \citenamefont
  {S\'{e}n\'{e}chal}}]{tremblayR}%
  \BibitemOpen
  \bibfield  {author} {\bibinfo {author} {\bibfnamefont {A.-M.~S.}\
  \bibnamefont {Tremblay}}, \bibinfo {author} {\bibfnamefont {B.}~\bibnamefont
  {Kyung}},\ and\ \bibinfo {author} {\bibfnamefont {D.}~\bibnamefont
  {S\'{e}n\'{e}chal}},\ }\bibfield  {title} {\bibinfo {title} {{Pseudogap and
  high-temperature superconductivity from weak to strong coupling. Towards a
  quantitative theory}},\ }\href {https://doi.org/10.1063/1.2199446} {\bibfield
   {journal} {\bibinfo  {journal} {Low Temp. Phys.}\ }\textbf {\bibinfo
  {volume} {32}},\ \bibinfo {pages} {424} (\bibinfo {year} {2006})}\BibitemShut
  {NoStop}%
\bibitem [{\citenamefont {Qin}\ \emph {et~al.}(2022)\citenamefont {Qin},
  \citenamefont {Sch\"{a}fer}, \citenamefont {Andergassen}, \citenamefont
  {Corboz},\ and\ \citenamefont {Gull}}]{QinAnnuRev2022}%
  \BibitemOpen
  \bibfield  {author} {\bibinfo {author} {\bibfnamefont {M.}~\bibnamefont
  {Qin}}, \bibinfo {author} {\bibfnamefont {T.}~\bibnamefont {Sch\"{a}fer}},
  \bibinfo {author} {\bibfnamefont {S.}~\bibnamefont {Andergassen}}, \bibinfo
  {author} {\bibfnamefont {P.}~\bibnamefont {Corboz}},\ and\ \bibinfo {author}
  {\bibfnamefont {E.}~\bibnamefont {Gull}},\ }\bibfield  {title} {\bibinfo
  {title} {{The Hubbard Model: A Computational Perspective}},\ }\href
  {https://doi.org/10.1146/annurev-conmatphys-090921-033948} {\bibfield
  {journal} {\bibinfo  {journal} {Annual Review of Condensed Matter Physics}\
  }\textbf {\bibinfo {volume} {13}},\ \bibinfo {pages} {275} (\bibinfo {year}
  {2022})}\BibitemShut {NoStop}%
\bibitem [{\citenamefont {Kotliar}\ \emph {et~al.}(2006)\citenamefont
  {Kotliar}, \citenamefont {Savrasov}, \citenamefont {Haule}, \citenamefont
  {Oudovenko}, \citenamefont {Parcollet},\ and\ \citenamefont
  {Marianetti}}]{kotliarRMP}%
  \BibitemOpen
  \bibfield  {author} {\bibinfo {author} {\bibfnamefont {G.}~\bibnamefont
  {Kotliar}}, \bibinfo {author} {\bibfnamefont {S.~Y.}\ \bibnamefont
  {Savrasov}}, \bibinfo {author} {\bibfnamefont {K.}~\bibnamefont {Haule}},
  \bibinfo {author} {\bibfnamefont {V.~S.}\ \bibnamefont {Oudovenko}}, \bibinfo
  {author} {\bibfnamefont {O.}~\bibnamefont {Parcollet}},\ and\ \bibinfo
  {author} {\bibfnamefont {C.~A.}\ \bibnamefont {Marianetti}},\ }\bibfield
  {title} {\bibinfo {title} {{Electronic structure calculations with dynamical
  mean-field theory}},\ }\href {https://doi.org/10.1103/RevModPhys.78.865}
  {\bibfield  {journal} {\bibinfo  {journal} {Rev. Mod. Phys.}\ }\textbf
  {\bibinfo {volume} {78}},\ \bibinfo {eid} {865} (\bibinfo {year}
  {2006})}\BibitemShut {NoStop}%
\bibitem [{\citenamefont {Georges}\ \emph {et~al.}(1996)\citenamefont
  {Georges}, \citenamefont {Kotliar}, \citenamefont {Krauth},\ and\
  \citenamefont {Rozenberg}}]{rmp}%
  \BibitemOpen
  \bibfield  {author} {\bibinfo {author} {\bibfnamefont {A.}~\bibnamefont
  {Georges}}, \bibinfo {author} {\bibfnamefont {G.}~\bibnamefont {Kotliar}},
  \bibinfo {author} {\bibfnamefont {W.}~\bibnamefont {Krauth}},\ and\ \bibinfo
  {author} {\bibfnamefont {M.~J.}\ \bibnamefont {Rozenberg}},\ }\bibfield
  {title} {\bibinfo {title} {{Dynamical mean-field theory of strongly
  correlated fermion systems and the limit of infinite dimensions}},\ }\href
  {https://doi.org/10.1103/RevModPhys.68.13} {\bibfield  {journal} {\bibinfo
  {journal} {Rev. Mod. Phys.}\ }\textbf {\bibinfo {volume} {68}},\ \bibinfo
  {pages} {13} (\bibinfo {year} {1996})}\BibitemShut {NoStop}%
\bibitem [{\citenamefont {Maier}\ \emph {et~al.}(2008)\citenamefont {Maier},
  \citenamefont {Poilblanc},\ and\ \citenamefont {Scalapino}}]{maierPRL2008}%
  \BibitemOpen
  \bibfield  {author} {\bibinfo {author} {\bibfnamefont {T.~A.}\ \bibnamefont
  {Maier}}, \bibinfo {author} {\bibfnamefont {D.}~\bibnamefont {Poilblanc}},\
  and\ \bibinfo {author} {\bibfnamefont {D.~J.}\ \bibnamefont {Scalapino}},\
  }\bibfield  {title} {\bibinfo {title} {{Dynamics of the Pairing Interaction
  in the Hubbard and $t\mathrm{\text{\ensuremath{-}}}J$ Models of
  High-Temperature Superconductors}},\ }\href
  {https://doi.org/10.1103/PhysRevLett.100.237001} {\bibfield  {journal}
  {\bibinfo  {journal} {Phys. Rev. Lett.}\ }\textbf {\bibinfo {volume} {100}},\
  \bibinfo {pages} {237001} (\bibinfo {year} {2008})}\BibitemShut {NoStop}%
\bibitem [{\citenamefont {Kyung}\ \emph {et~al.}(2009)\citenamefont {Kyung},
  \citenamefont {S\'{e}n\'{e}chal},\ and\ \citenamefont
  {Tremblay}}]{Kyung:2009}%
  \BibitemOpen
  \bibfield  {author} {\bibinfo {author} {\bibfnamefont {B.}~\bibnamefont
  {Kyung}}, \bibinfo {author} {\bibfnamefont {D.}~\bibnamefont
  {S\'{e}n\'{e}chal}},\ and\ \bibinfo {author} {\bibfnamefont {A.-M.~S.}\
  \bibnamefont {Tremblay}},\ }\bibfield  {title} {\bibinfo {title} {Pairing
  dynamics in strongly correlated superconductivity},\ }\href
  {https://doi.org/10.1103/PhysRevB.80.205109} {\bibfield  {journal} {\bibinfo
  {journal} {Physical Review B (Condensed Matter and Materials Physics)}\
  }\textbf {\bibinfo {volume} {80}},\ \bibinfo {eid} {205109} (\bibinfo {year}
  {2009})}\BibitemShut {NoStop}%
\bibitem [{\citenamefont {Civelli}(2009)}]{civelli1}%
  \BibitemOpen
  \bibfield  {author} {\bibinfo {author} {\bibfnamefont {M.}~\bibnamefont
  {Civelli}},\ }\bibfield  {title} {\bibinfo {title} {Evolution of the
  dynamical pairing across the phase diagram of a strongly correlated
  high-temperature superconductor},\ }\href
  {https://doi.org/10.1103/PhysRevLett.103.136402} {\bibfield  {journal}
  {\bibinfo  {journal} {Phys. Rev. Lett.}\ }\textbf {\bibinfo {volume} {103}},\
  \bibinfo {pages} {136402} (\bibinfo {year} {2009})}\BibitemShut {NoStop}%
\bibitem [{\citenamefont {S\'en\'echal}\ \emph {et~al.}(2013)\citenamefont
  {S\'en\'echal}, \citenamefont {Day}, \citenamefont {Bouliane},\ and\
  \citenamefont {Tremblay}}]{senechalPRB2013}%
  \BibitemOpen
  \bibfield  {author} {\bibinfo {author} {\bibfnamefont {D.}~\bibnamefont
  {S\'en\'echal}}, \bibinfo {author} {\bibfnamefont {A.~G.~R.}\ \bibnamefont
  {Day}}, \bibinfo {author} {\bibfnamefont {V.}~\bibnamefont {Bouliane}},\ and\
  \bibinfo {author} {\bibfnamefont {A.-M.~S.}\ \bibnamefont {Tremblay}},\
  }\bibfield  {title} {\bibinfo {title} {Resilience of $d$-wave
  superconductivity to nearest-neighbor repulsion},\ }\href
  {https://doi.org/10.1103/PhysRevB.87.075123} {\bibfield  {journal} {\bibinfo
  {journal} {Phys. Rev. B}\ }\textbf {\bibinfo {volume} {87}},\ \bibinfo
  {pages} {075123} (\bibinfo {year} {2013})}\BibitemShut {NoStop}%
\bibitem [{\citenamefont {Gull}\ and\ \citenamefont
  {Millis}(2014)}]{Gull:PRB2014}%
  \BibitemOpen
  \bibfield  {author} {\bibinfo {author} {\bibfnamefont {E.}~\bibnamefont
  {Gull}}\ and\ \bibinfo {author} {\bibfnamefont {A.~J.}\ \bibnamefont
  {Millis}},\ }\bibfield  {title} {\bibinfo {title} {{Pairing glue in the
  two-dimensional Hubbard model}},\ }\href
  {https://doi.org/10.1103/PhysRevB.90.041110} {\bibfield  {journal} {\bibinfo
  {journal} {Phys. Rev. B}\ }\textbf {\bibinfo {volume} {90}},\ \bibinfo
  {pages} {041110} (\bibinfo {year} {2014})}\BibitemShut {NoStop}%
\bibitem [{\citenamefont {Reymbaut}\ \emph {et~al.}(2016)\citenamefont
  {Reymbaut}, \citenamefont {Charlebois}, \citenamefont {Asiani}, \citenamefont
  {Fratino}, \citenamefont {S\'emon}, \citenamefont {Sordi},\ and\
  \citenamefont {Tremblay}}]{reymbautPRB2016}%
  \BibitemOpen
  \bibfield  {author} {\bibinfo {author} {\bibfnamefont {A.}~\bibnamefont
  {Reymbaut}}, \bibinfo {author} {\bibfnamefont {M.}~\bibnamefont
  {Charlebois}}, \bibinfo {author} {\bibfnamefont {M.~F.}\ \bibnamefont
  {Asiani}}, \bibinfo {author} {\bibfnamefont {L.}~\bibnamefont {Fratino}},
  \bibinfo {author} {\bibfnamefont {P.}~\bibnamefont {S\'emon}}, \bibinfo
  {author} {\bibfnamefont {G.}~\bibnamefont {Sordi}},\ and\ \bibinfo {author}
  {\bibfnamefont {A.-M.~S.}\ \bibnamefont {Tremblay}},\ }\bibfield  {title}
  {\bibinfo {title} {{Antagonistic effects of nearest-neighbor repulsion on the
  superconducting pairing dynamics in the doped Mott insulator regime}},\
  }\href {https://doi.org/10.1103/PhysRevB.94.155146} {\bibfield  {journal}
  {\bibinfo  {journal} {Phys. Rev. B}\ }\textbf {\bibinfo {volume} {94}},\
  \bibinfo {pages} {155146} (\bibinfo {year} {2016})}\BibitemShut {NoStop}%
\bibitem [{\citenamefont {{Dong}}\ \emph
  {et~al.}(2022{\natexlab{a}})\citenamefont {{Dong}}, \citenamefont {{Del Re}},
  \citenamefont {{Toschi}},\ and\ \citenamefont {{Gull}}}]{DongPNAS2022}%
  \BibitemOpen
  \bibfield  {author} {\bibinfo {author} {\bibfnamefont {X.}~\bibnamefont
  {{Dong}}}, \bibinfo {author} {\bibfnamefont {L.}~\bibnamefont {{Del Re}}},
  \bibinfo {author} {\bibfnamefont {A.}~\bibnamefont {{Toschi}}},\ and\
  \bibinfo {author} {\bibfnamefont {E.}~\bibnamefont {{Gull}}},\ }\bibfield
  {title} {\bibinfo {title} {{Mechanism of superconductivity in the Hubbard
  model at intermediate interaction strength}},\ }\href
  {https://doi.org/10.1073/pnas.2205048119} {\bibfield  {journal} {\bibinfo
  {journal} {Proceedings of the National Academy of Science}\ }\textbf
  {\bibinfo {volume} {119}},\ \bibinfo {eid} {e2205048119} (\bibinfo {year}
  {2022}{\natexlab{a}})}\BibitemShut {NoStop}%
\bibitem [{\citenamefont {{Dong}}\ \emph
  {et~al.}(2022{\natexlab{b}})\citenamefont {{Dong}}, \citenamefont {{Gull}},\
  and\ \citenamefont {{Millis}}}]{DongNatPhys2022}%
  \BibitemOpen
  \bibfield  {author} {\bibinfo {author} {\bibfnamefont {X.}~\bibnamefont
  {{Dong}}}, \bibinfo {author} {\bibfnamefont {E.}~\bibnamefont {{Gull}}},\
  and\ \bibinfo {author} {\bibfnamefont {A.~J.}\ \bibnamefont {{Millis}}},\
  }\bibfield  {title} {\bibinfo {title} {{Quantifying the role of
  antiferromagnetic fluctuations in the superconductivity of the doped Hubbard
  model}},\ }\href {https://doi.org/10.1038/s41567-022-01710-z} {\bibfield
  {journal} {\bibinfo  {journal} {Nature Physics}\ }\textbf {\bibinfo {volume}
  {18}},\ \bibinfo {pages} {1293} (\bibinfo {year}
  {2022}{\natexlab{b}})}\BibitemShut {NoStop}%
\bibitem [{\citenamefont {Haule}(2007)}]{hauleCTQMC}%
  \BibitemOpen
  \bibfield  {author} {\bibinfo {author} {\bibfnamefont {K.}~\bibnamefont
  {Haule}},\ }\bibfield  {title} {\bibinfo {title} {{Quantum Monte Carlo
  impurity solver for cluster dynamical mean-field theory and electronic
  structure calculations with adjustable cluster base}},\ }\href
  {https://doi.org/10.1103/PhysRevB.75.155113} {\bibfield  {journal} {\bibinfo
  {journal} {Phys. Rev. B}\ }\textbf {\bibinfo {volume} {75}},\ \bibinfo {eid}
  {155113} (\bibinfo {year} {2007})}\BibitemShut {NoStop}%
\bibitem [{\citenamefont {Werner}\ \emph {et~al.}(2006)\citenamefont {Werner},
  \citenamefont {Comanac}, \citenamefont {de~Medici}, \citenamefont {Troyer},\
  and\ \citenamefont {Millis}}]{Werner:2006}%
  \BibitemOpen
  \bibfield  {author} {\bibinfo {author} {\bibfnamefont {P.}~\bibnamefont
  {Werner}}, \bibinfo {author} {\bibfnamefont {A.}~\bibnamefont {Comanac}},
  \bibinfo {author} {\bibfnamefont {L.}~\bibnamefont {de~Medici}}, \bibinfo
  {author} {\bibfnamefont {M.}~\bibnamefont {Troyer}},\ and\ \bibinfo {author}
  {\bibfnamefont {A.~J.}\ \bibnamefont {Millis}},\ }\bibfield  {title}
  {\bibinfo {title} {Continuous-time solver for quantum impurity models},\
  }\href {https://doi.org/10.1103/PhysRevLett.97.076405} {\bibfield  {journal}
  {\bibinfo  {journal} {Phys. Rev. Lett.}\ }\textbf {\bibinfo {volume} {97}},\
  \bibinfo {pages} {076405} (\bibinfo {year} {2006})}\BibitemShut {NoStop}%
\bibitem [{\citenamefont {Gull}\ \emph {et~al.}(2011)\citenamefont {Gull},
  \citenamefont {Millis}, \citenamefont {Lichtenstein}, \citenamefont
  {Rubtsov}, \citenamefont {Troyer},\ and\ \citenamefont {Werner}}]{millisRMP}%
  \BibitemOpen
  \bibfield  {author} {\bibinfo {author} {\bibfnamefont {E.}~\bibnamefont
  {Gull}}, \bibinfo {author} {\bibfnamefont {A.~J.}\ \bibnamefont {Millis}},
  \bibinfo {author} {\bibfnamefont {A.~I.}\ \bibnamefont {Lichtenstein}},
  \bibinfo {author} {\bibfnamefont {A.~N.}\ \bibnamefont {Rubtsov}}, \bibinfo
  {author} {\bibfnamefont {M.}~\bibnamefont {Troyer}},\ and\ \bibinfo {author}
  {\bibfnamefont {P.}~\bibnamefont {Werner}},\ }\bibfield  {title} {\bibinfo
  {title} {{Continuous-time Monte~Carlo methods for quantum impurity models}},\
  }\href {https://doi.org/10.1103/RevModPhys.83.349} {\bibfield  {journal}
  {\bibinfo  {journal} {Rev. Mod. Phys.}\ }\textbf {\bibinfo {volume} {83}},\
  \bibinfo {pages} {349} (\bibinfo {year} {2011})}\BibitemShut {NoStop}%
\bibitem [{\citenamefont {S\'emon}\ \emph
  {et~al.}(2014{\natexlab{a}})\citenamefont {S\'emon}, \citenamefont {Yee},
  \citenamefont {Haule},\ and\ \citenamefont {Tremblay}}]{patrickSkipList}%
  \BibitemOpen
  \bibfield  {author} {\bibinfo {author} {\bibfnamefont {P.}~\bibnamefont
  {S\'emon}}, \bibinfo {author} {\bibfnamefont {C.-H.}\ \bibnamefont {Yee}},
  \bibinfo {author} {\bibfnamefont {K.}~\bibnamefont {Haule}},\ and\ \bibinfo
  {author} {\bibfnamefont {A.-M.~S.}\ \bibnamefont {Tremblay}},\ }\bibfield
  {title} {\bibinfo {title} {{Lazy skip-lists: An algorithm for fast
  hybridization-expansion quantum Monte Carlo}},\ }\href
  {https://doi.org/10.1103/PhysRevB.90.075149} {\bibfield  {journal} {\bibinfo
  {journal} {Phys. Rev. B}\ }\textbf {\bibinfo {volume} {90}},\ \bibinfo
  {pages} {075149} (\bibinfo {year} {2014}{\natexlab{a}})}\BibitemShut
  {NoStop}%
\bibitem [{\citenamefont {S\'emon}\ \emph
  {et~al.}(2014{\natexlab{b}})\citenamefont {S\'emon}, \citenamefont {Sordi},\
  and\ \citenamefont {Tremblay}}]{patrickERG}%
  \BibitemOpen
  \bibfield  {author} {\bibinfo {author} {\bibfnamefont {P.}~\bibnamefont
  {S\'emon}}, \bibinfo {author} {\bibfnamefont {G.}~\bibnamefont {Sordi}},\
  and\ \bibinfo {author} {\bibfnamefont {A.-M.~S.}\ \bibnamefont {Tremblay}},\
  }\bibfield  {title} {\bibinfo {title} {{Ergodicity of the
  hybridization-expansion Monte Carlo algorithm for broken-symmetry states}},\
  }\href {https://doi.org/10.1103/PhysRevB.89.165113} {\bibfield  {journal}
  {\bibinfo  {journal} {Phys. Rev. B}\ }\textbf {\bibinfo {volume} {89}},\
  \bibinfo {pages} {165113} (\bibinfo {year} {2014}{\natexlab{b}})}\BibitemShut
  {NoStop}%
\bibitem [{\citenamefont {Sakai}(2023)}]{Sakai:JPSJ2023}%
  \BibitemOpen
  \bibfield  {author} {\bibinfo {author} {\bibfnamefont {S.}~\bibnamefont
  {Sakai}},\ }\bibfield  {title} {\bibinfo {title} {Nonperturbative
  calculations for spectroscopic properties of cuprate high-temperature
  superconductors},\ }\href {https://doi.org/10.7566/JPSJ.92.092001} {\bibfield
   {journal} {\bibinfo  {journal} {Journal of the Physical Society of Japan}\
  }\textbf {\bibinfo {volume} {92}},\ \bibinfo {pages} {092001} (\bibinfo
  {year} {2023})}\BibitemShut {NoStop}%
\bibitem [{\citenamefont {Liu}\ \emph {et~al.}(2025)\citenamefont {Liu},
  \citenamefont {Yao},\ and\ \citenamefont {Wu}}]{Liu:CPL2025}%
  \BibitemOpen
  \bibfield  {author} {\bibinfo {author} {\bibfnamefont {J.}~\bibnamefont
  {Liu}}, \bibinfo {author} {\bibfnamefont {D.-X.}\ \bibnamefont {Yao}},\ and\
  \bibinfo {author} {\bibfnamefont {W.}~\bibnamefont {Wu}},\ }\bibfield
  {title} {\bibinfo {title} {{Interplay between the Pseudogap and
  Superconductivity in Doped Mott Insulators: a Cluster Dynamical Mean-Field
  Theory Study}},\ }\href {https://doi.org/10.1088/0256-307X/42/8/080711}
  {\bibfield  {journal} {\bibinfo  {journal} {Chinese Physics Letters}\
  }\textbf {\bibinfo {volume} {42}},\ \bibinfo {pages} {080711} (\bibinfo
  {year} {2025})}\BibitemShut {NoStop}%
\bibitem [{\citenamefont {Kancharla}\ \emph {et~al.}(2008)\citenamefont
  {Kancharla}, \citenamefont {Kyung}, \citenamefont {S\'en\'echal},
  \citenamefont {Civelli}, \citenamefont {Capone}, \citenamefont {Kotliar},\
  and\ \citenamefont {Tremblay}}]{kancharla}%
  \BibitemOpen
  \bibfield  {author} {\bibinfo {author} {\bibfnamefont {S.~S.}\ \bibnamefont
  {Kancharla}}, \bibinfo {author} {\bibfnamefont {B.}~\bibnamefont {Kyung}},
  \bibinfo {author} {\bibfnamefont {D.}~\bibnamefont {S\'en\'echal}}, \bibinfo
  {author} {\bibfnamefont {M.}~\bibnamefont {Civelli}}, \bibinfo {author}
  {\bibfnamefont {M.}~\bibnamefont {Capone}}, \bibinfo {author} {\bibfnamefont
  {G.}~\bibnamefont {Kotliar}},\ and\ \bibinfo {author} {\bibfnamefont
  {A.-M.~S.}\ \bibnamefont {Tremblay}},\ }\bibfield  {title} {\bibinfo {title}
  {{Anomalous superconductivity and its competition with antiferromagnetism in
  doped Mott insulators}},\ }\href {https://doi.org/10.1103/PhysRevB.77.184516}
  {\bibfield  {journal} {\bibinfo  {journal} {Phys. Rev. B}\ }\textbf {\bibinfo
  {volume} {77}},\ \bibinfo {pages} {184516} (\bibinfo {year}
  {2008})}\BibitemShut {NoStop}%
\bibitem [{\citenamefont {Haule}\ and\ \citenamefont
  {Kotliar}(2007)}]{hauleDOPING}%
  \BibitemOpen
  \bibfield  {author} {\bibinfo {author} {\bibfnamefont {K.}~\bibnamefont
  {Haule}}\ and\ \bibinfo {author} {\bibfnamefont {G.}~\bibnamefont
  {Kotliar}},\ }\bibfield  {title} {\bibinfo {title} {{Strongly correlated
  superconductivity: A plaquette dynamical mean-field theory study}},\ }\href
  {https://doi.org/10.1103/PhysRevB.76.104509} {\bibfield  {journal} {\bibinfo
  {journal} {Phys. Rev. B}\ }\textbf {\bibinfo {volume} {76}},\ \bibinfo {eid}
  {104509} (\bibinfo {year} {2007})}\BibitemShut {NoStop}%
\bibitem [{\citenamefont {Sordi}\ \emph
  {et~al.}(2012{\natexlab{a}})\citenamefont {Sordi}, \citenamefont {S\'emon},
  \citenamefont {Haule},\ and\ \citenamefont {Tremblay}}]{sshtSC}%
  \BibitemOpen
  \bibfield  {author} {\bibinfo {author} {\bibfnamefont {G.}~\bibnamefont
  {Sordi}}, \bibinfo {author} {\bibfnamefont {P.}~\bibnamefont {S\'emon}},
  \bibinfo {author} {\bibfnamefont {K.}~\bibnamefont {Haule}},\ and\ \bibinfo
  {author} {\bibfnamefont {A.-M.~S.}\ \bibnamefont {Tremblay}},\ }\bibfield
  {title} {\bibinfo {title} {{Strong Coupling Superconductivity, Pseudogap, and
  Mott Transition}},\ }\href {https://doi.org/10.1103/PhysRevLett.108.216401}
  {\bibfield  {journal} {\bibinfo  {journal} {Phys. Rev. Lett.}\ }\textbf
  {\bibinfo {volume} {108}},\ \bibinfo {pages} {216401} (\bibinfo {year}
  {2012}{\natexlab{a}})}\BibitemShut {NoStop}%
\bibitem [{\citenamefont {Fratino}\ \emph {et~al.}(2016)\citenamefont
  {Fratino}, \citenamefont {S\'emon}, \citenamefont {Sordi},\ and\
  \citenamefont {Tremblay}}]{LorenzoSC}%
  \BibitemOpen
  \bibfield  {author} {\bibinfo {author} {\bibfnamefont {L.}~\bibnamefont
  {Fratino}}, \bibinfo {author} {\bibfnamefont {P.}~\bibnamefont {S\'emon}},
  \bibinfo {author} {\bibfnamefont {G.}~\bibnamefont {Sordi}},\ and\ \bibinfo
  {author} {\bibfnamefont {A.-M.~S.}\ \bibnamefont {Tremblay}},\ }\bibfield
  {title} {\bibinfo {title} {{An organizing principle for two-dimensional
  strongly correlated superconductivity}},\ }\href
  {https://doi.org/10.1038/srep22715} {\bibfield  {journal} {\bibinfo
  {journal} {Sci. Rep.}\ }\textbf {\bibinfo {volume} {6}},\ \bibinfo {pages}
  {22715} (\bibinfo {year} {2016})}\BibitemShut {NoStop}%
\bibitem [{\citenamefont {H\'ebert}\ \emph {et~al.}(2015)\citenamefont
  {H\'ebert}, \citenamefont {S\'emon},\ and\ \citenamefont
  {Tremblay}}]{Hebert:2015}%
  \BibitemOpen
  \bibfield  {author} {\bibinfo {author} {\bibfnamefont {C.-D.}\ \bibnamefont
  {H\'ebert}}, \bibinfo {author} {\bibfnamefont {P.}~\bibnamefont {S\'emon}},\
  and\ \bibinfo {author} {\bibfnamefont {A.-M.~S.}\ \bibnamefont {Tremblay}},\
  }\bibfield  {title} {\bibinfo {title} {Superconducting dome in doped
  quasi-two-dimensional organic mott insulators: A paradigm for strongly
  correlated superconductivity},\ }\href
  {https://doi.org/10.1103/PhysRevB.92.195112} {\bibfield  {journal} {\bibinfo
  {journal} {Phys. Rev. B}\ }\textbf {\bibinfo {volume} {92}},\ \bibinfo
  {pages} {195112} (\bibinfo {year} {2015})}\BibitemShut {NoStop}%
\bibitem [{\citenamefont {S\'en\'echal}\ \emph {et~al.}(2005)\citenamefont
  {S\'en\'echal}, \citenamefont {Lavertu}, \citenamefont {Marois},\ and\
  \citenamefont {Tremblay}}]{senechalAFSC2005}%
  \BibitemOpen
  \bibfield  {author} {\bibinfo {author} {\bibfnamefont {D.}~\bibnamefont
  {S\'en\'echal}}, \bibinfo {author} {\bibfnamefont {P.-L.}\ \bibnamefont
  {Lavertu}}, \bibinfo {author} {\bibfnamefont {M.-A.}\ \bibnamefont
  {Marois}},\ and\ \bibinfo {author} {\bibfnamefont {A.-M.~S.}\ \bibnamefont
  {Tremblay}},\ }\bibfield  {title} {\bibinfo {title} {Competition between
  antiferromagnetism and superconductivity in high-${T}_{c}$ cuprates},\ }\href
  {https://doi.org/10.1103/PhysRevLett.94.156404} {\bibfield  {journal}
  {\bibinfo  {journal} {Phys. Rev. Lett.}\ }\textbf {\bibinfo {volume} {94}},\
  \bibinfo {pages} {156404} (\bibinfo {year} {2005})}\BibitemShut {NoStop}%
\bibitem [{\citenamefont {Civelli}\ \emph {et~al.}(2008)\citenamefont
  {Civelli}, \citenamefont {Capone}, \citenamefont {Georges}, \citenamefont
  {Haule}, \citenamefont {Parcollet}, \citenamefont {Stanescu},\ and\
  \citenamefont {Kotliar}}]{Civelli:PRL2008}%
  \BibitemOpen
  \bibfield  {author} {\bibinfo {author} {\bibfnamefont {M.}~\bibnamefont
  {Civelli}}, \bibinfo {author} {\bibfnamefont {M.}~\bibnamefont {Capone}},
  \bibinfo {author} {\bibfnamefont {A.}~\bibnamefont {Georges}}, \bibinfo
  {author} {\bibfnamefont {K.}~\bibnamefont {Haule}}, \bibinfo {author}
  {\bibfnamefont {O.}~\bibnamefont {Parcollet}}, \bibinfo {author}
  {\bibfnamefont {T.~D.}\ \bibnamefont {Stanescu}},\ and\ \bibinfo {author}
  {\bibfnamefont {G.}~\bibnamefont {Kotliar}},\ }\bibfield  {title} {\bibinfo
  {title} {{Nodal-Antinodal Dichotomy and the Two Gaps of a Superconducting
  Doped Mott Insulator}},\ }\href
  {https://doi.org/10.1103/PhysRevLett.100.046402} {\bibfield  {journal}
  {\bibinfo  {journal} {Phys. Rev. Lett.}\ }\textbf {\bibinfo {volume} {100}},\
  \bibinfo {pages} {046402} (\bibinfo {year} {2008})}\BibitemShut {NoStop}%
\bibitem [{\citenamefont {Walsh}\ \emph {et~al.}(2023)\citenamefont {Walsh},
  \citenamefont {Charlebois}, \citenamefont {S\'emon}, \citenamefont
  {Tremblay},\ and\ \citenamefont {Sordi}}]{Walsh:PRB2023}%
  \BibitemOpen
  \bibfield  {author} {\bibinfo {author} {\bibfnamefont {C.}~\bibnamefont
  {Walsh}}, \bibinfo {author} {\bibfnamefont {M.}~\bibnamefont {Charlebois}},
  \bibinfo {author} {\bibfnamefont {P.}~\bibnamefont {S\'emon}}, \bibinfo
  {author} {\bibfnamefont {A.-M.~S.}\ \bibnamefont {Tremblay}},\ and\ \bibinfo
  {author} {\bibfnamefont {G.}~\bibnamefont {Sordi}},\ }\bibfield  {title}
  {\bibinfo {title} {{Superconductivity in the two-dimensional Hubbard model
  with cellular dynamical mean-field theory: A quantum impurity model
  analysis}},\ }\href {https://doi.org/10.1103/PhysRevB.108.075163} {\bibfield
  {journal} {\bibinfo  {journal} {Phys. Rev. B}\ }\textbf {\bibinfo {volume}
  {108}},\ \bibinfo {pages} {075163} (\bibinfo {year} {2023})}\BibitemShut
  {NoStop}%
\bibitem [{\citenamefont {Carbone}\ \emph {et~al.}(2006)\citenamefont
  {Carbone}, \citenamefont {Kuzmenko}, \citenamefont {Molegraaf}, \citenamefont
  {van Heumen}, \citenamefont {Lukovac}, \citenamefont {Marsiglio},
  \citenamefont {van~der Marel}, \citenamefont {Haule}, \citenamefont
  {Kotliar}, \citenamefont {Berger}, \citenamefont {Courjault}, \citenamefont
  {Kes},\ and\ \citenamefont {Li}}]{carbone2006}%
  \BibitemOpen
  \bibfield  {author} {\bibinfo {author} {\bibfnamefont {F.}~\bibnamefont
  {Carbone}}, \bibinfo {author} {\bibfnamefont {A.~B.}\ \bibnamefont
  {Kuzmenko}}, \bibinfo {author} {\bibfnamefont {H.~J.~A.}\ \bibnamefont
  {Molegraaf}}, \bibinfo {author} {\bibfnamefont {E.}~\bibnamefont {van
  Heumen}}, \bibinfo {author} {\bibfnamefont {V.}~\bibnamefont {Lukovac}},
  \bibinfo {author} {\bibfnamefont {F.}~\bibnamefont {Marsiglio}}, \bibinfo
  {author} {\bibfnamefont {D.}~\bibnamefont {van~der Marel}}, \bibinfo {author}
  {\bibfnamefont {K.}~\bibnamefont {Haule}}, \bibinfo {author} {\bibfnamefont
  {G.}~\bibnamefont {Kotliar}}, \bibinfo {author} {\bibfnamefont
  {H.}~\bibnamefont {Berger}}, \bibinfo {author} {\bibfnamefont
  {S.}~\bibnamefont {Courjault}}, \bibinfo {author} {\bibfnamefont {P.~H.}\
  \bibnamefont {Kes}},\ and\ \bibinfo {author} {\bibfnamefont {M.}~\bibnamefont
  {Li}},\ }\bibfield  {title} {\bibinfo {title} {Doping dependence of the
  redistribution of optical spectral weight in
  ${\mathrm{bi}}_{2}{\mathrm{sr}}_{2}{\mathrm{cacu}}_{2}{\mathrm{o}}_{8+\ensuremath{\delta}}$},\
  }\href {https://doi.org/10.1103/PhysRevB.74.064510} {\bibfield  {journal}
  {\bibinfo  {journal} {Phys. Rev. B}\ }\textbf {\bibinfo {volume} {74}},\
  \bibinfo {pages} {064510} (\bibinfo {year} {2006})}\BibitemShut {NoStop}%
\bibitem [{\citenamefont {Walsh}\ \emph {et~al.}(2021)\citenamefont {Walsh},
  \citenamefont {Charlebois}, \citenamefont {Sémon}, \citenamefont {Sordi},\
  and\ \citenamefont {Tremblay}}]{CaitlinPNAS2021}%
  \BibitemOpen
  \bibfield  {author} {\bibinfo {author} {\bibfnamefont {C.}~\bibnamefont
  {Walsh}}, \bibinfo {author} {\bibfnamefont {M.}~\bibnamefont {Charlebois}},
  \bibinfo {author} {\bibfnamefont {P.}~\bibnamefont {Sémon}}, \bibinfo
  {author} {\bibfnamefont {G.}~\bibnamefont {Sordi}},\ and\ \bibinfo {author}
  {\bibfnamefont {A.-M.~S.}\ \bibnamefont {Tremblay}},\ }\bibfield  {title}
  {\bibinfo {title} {{Information-theoretic measures of superconductivity in a
  two-dimensional doped Mott insulator}},\ }\href
  {https://doi.org/10.1073/pnas.2104114118} {\bibfield  {journal} {\bibinfo
  {journal} {Proceedings of the National Academy of Sciences}\ }\textbf
  {\bibinfo {volume} {118}},\ \bibinfo {pages} {e2104114118} (\bibinfo {year}
  {2021})}\BibitemShut {NoStop}%
\bibitem [{\citenamefont {Bergeron}\ and\ \citenamefont
  {Tremblay}(2016)}]{DominicMEM}%
  \BibitemOpen
  \bibfield  {author} {\bibinfo {author} {\bibfnamefont {D.}~\bibnamefont
  {Bergeron}}\ and\ \bibinfo {author} {\bibfnamefont {A.-M.~S.}\ \bibnamefont
  {Tremblay}},\ }\bibfield  {title} {\bibinfo {title} {Algorithms for optimized
  maximum entropy and diagnostic tools for analytic continuation},\ }\href
  {https://doi.org/10.1103/PhysRevE.94.023303} {\bibfield  {journal} {\bibinfo
  {journal} {Phys. Rev. E}\ }\textbf {\bibinfo {volume} {94}},\ \bibinfo
  {pages} {023303} (\bibinfo {year} {2016})}\BibitemShut {NoStop}%
\bibitem [{\citenamefont {Reymbaut}\ \emph {et~al.}(2015)\citenamefont
  {Reymbaut}, \citenamefont {Bergeron},\ and\ \citenamefont
  {Tremblay}}]{Alexis_PRB2015}%
  \BibitemOpen
  \bibfield  {author} {\bibinfo {author} {\bibfnamefont {A.}~\bibnamefont
  {Reymbaut}}, \bibinfo {author} {\bibfnamefont {D.}~\bibnamefont {Bergeron}},\
  and\ \bibinfo {author} {\bibfnamefont {A.-M.~S.}\ \bibnamefont {Tremblay}},\
  }\bibfield  {title} {\bibinfo {title} {{Maximum entropy analytic continuation
  for spectral functions with nonpositive spectral weight}},\ }\href
  {https://doi.org/10.1103/PhysRevB.92.060509} {\bibfield  {journal} {\bibinfo
  {journal} {Phys. Rev. B}\ }\textbf {\bibinfo {volume} {92}},\ \bibinfo
  {pages} {060509} (\bibinfo {year} {2015})}\BibitemShut {NoStop}%
\bibitem [{Sup()}]{SupplementalMaterial}%
  \BibitemOpen
  \href@noop {} {}\bibinfo {note} {See Supplemental Material for the
  calculation of $A_{\rm an}(\omega)$ by performing only a single analytical
  continuation, the consistency checks for $A_{\rm an}(\omega)$, extended data
  for $A_{\rm nor}(\omega)$, $A_{\rm an}(\omega)$, $I_{F}(\omega)$, and the
  computation of $A_{\rm an}(t)$ in the time domain.}\BibitemShut {Stop}%
\bibitem [{Note1()}]{Note1}%
  \BibitemOpen
  \bibinfo {note} {This neglects Kosterlitz-Thouless physics}\BibitemShut
  {NoStop}%
\bibitem [{\citenamefont {Walsh}\ \emph {et~al.}(2019)\citenamefont {Walsh},
  \citenamefont {S\'emon}, \citenamefont {Poulin}, \citenamefont {Sordi},\ and\
  \citenamefont {Tremblay}}]{CaitlinSb}%
  \BibitemOpen
  \bibfield  {author} {\bibinfo {author} {\bibfnamefont {C.}~\bibnamefont
  {Walsh}}, \bibinfo {author} {\bibfnamefont {P.}~\bibnamefont {S\'emon}},
  \bibinfo {author} {\bibfnamefont {D.}~\bibnamefont {Poulin}}, \bibinfo
  {author} {\bibfnamefont {G.}~\bibnamefont {Sordi}},\ and\ \bibinfo {author}
  {\bibfnamefont {A.-M.~S.}\ \bibnamefont {Tremblay}},\ }\bibfield  {title}
  {\bibinfo {title} {{Thermodynamic and information-theoretic description of
  the Mott transition in the two-dimensional Hubbard model}},\ }\href
  {https://doi.org/10.1103/PhysRevB.99.075122} {\bibfield  {journal} {\bibinfo
  {journal} {Phys. Rev. B}\ }\textbf {\bibinfo {volume} {99}},\ \bibinfo
  {pages} {075122} (\bibinfo {year} {2019})}\BibitemShut {NoStop}%
\bibitem [{\citenamefont {Paramekanti}\ \emph {et~al.}(2004)\citenamefont
  {Paramekanti}, \citenamefont {Randeria},\ and\ \citenamefont
  {Trivedi}}]{Paramekanti:2004}%
  \BibitemOpen
  \bibfield  {author} {\bibinfo {author} {\bibfnamefont {A.}~\bibnamefont
  {Paramekanti}}, \bibinfo {author} {\bibfnamefont {M.}~\bibnamefont
  {Randeria}},\ and\ \bibinfo {author} {\bibfnamefont {N.}~\bibnamefont
  {Trivedi}},\ }\bibfield  {title} {\bibinfo {title} {{High- $T_{c}$
  superconductors: A variational theory of the superconducting state}},\ }\href
  {https://doi.org/10.1103/PhysRevB.70.054504} {\bibfield  {journal} {\bibinfo
  {journal} {Phys. Rev. B}\ }\textbf {\bibinfo {volume} {70}},\ \bibinfo
  {pages} {054504} (\bibinfo {year} {2004})}\BibitemShut {NoStop}%
\bibitem [{\citenamefont {Gull}\ \emph {et~al.}(2013)\citenamefont {Gull},
  \citenamefont {Parcollet},\ and\ \citenamefont {Millis}}]{Gull:2013}%
  \BibitemOpen
  \bibfield  {author} {\bibinfo {author} {\bibfnamefont {E.}~\bibnamefont
  {Gull}}, \bibinfo {author} {\bibfnamefont {O.}~\bibnamefont {Parcollet}},\
  and\ \bibinfo {author} {\bibfnamefont {A.~J.}\ \bibnamefont {Millis}},\
  }\bibfield  {title} {\bibinfo {title} {{Superconductivity and the Pseudogap
  in the Two-Dimensional Hubbard Model}},\ }\href
  {https://doi.org/10.1103/PhysRevLett.110.216405} {\bibfield  {journal}
  {\bibinfo  {journal} {Phys. Rev. Lett.}\ }\textbf {\bibinfo {volume} {110}},\
  \bibinfo {pages} {216405} (\bibinfo {year} {2013})}\BibitemShut {NoStop}%
\bibitem [{\citenamefont {Sobota}\ \emph {et~al.}(2021)\citenamefont {Sobota},
  \citenamefont {He},\ and\ \citenamefont {Shen}}]{Sobota:RMP2021}%
  \BibitemOpen
  \bibfield  {author} {\bibinfo {author} {\bibfnamefont {J.~A.}\ \bibnamefont
  {Sobota}}, \bibinfo {author} {\bibfnamefont {Y.}~\bibnamefont {He}},\ and\
  \bibinfo {author} {\bibfnamefont {Z.-X.}\ \bibnamefont {Shen}},\ }\bibfield
  {title} {\bibinfo {title} {Angle-resolved photoemission studies of quantum
  materials},\ }\href {https://doi.org/10.1103/RevModPhys.93.025006} {\bibfield
   {journal} {\bibinfo  {journal} {Rev. Mod. Phys.}\ }\textbf {\bibinfo
  {volume} {93}},\ \bibinfo {pages} {025006} (\bibinfo {year}
  {2021})}\BibitemShut {NoStop}%
\bibitem [{\citenamefont {Sordi}\ \emph {et~al.}(2010)\citenamefont {Sordi},
  \citenamefont {Haule},\ and\ \citenamefont {Tremblay}}]{sht}%
  \BibitemOpen
  \bibfield  {author} {\bibinfo {author} {\bibfnamefont {G.}~\bibnamefont
  {Sordi}}, \bibinfo {author} {\bibfnamefont {K.}~\bibnamefont {Haule}},\ and\
  \bibinfo {author} {\bibfnamefont {A.-M.~S.}\ \bibnamefont {Tremblay}},\
  }\bibfield  {title} {\bibinfo {title} {{Finite Doping Signatures of the Mott
  Transition in the Two-Dimensional Hubbard Model}},\ }\href
  {https://doi.org/10.1103/PhysRevLett.104.226402} {\bibfield  {journal}
  {\bibinfo  {journal} {Phys. Rev. Lett.}\ }\textbf {\bibinfo {volume} {104}},\
  \bibinfo {pages} {226402} (\bibinfo {year} {2010})}\BibitemShut {NoStop}%
\bibitem [{\citenamefont {Sordi}\ \emph {et~al.}(2011)\citenamefont {Sordi},
  \citenamefont {Haule},\ and\ \citenamefont {Tremblay}}]{sht2}%
  \BibitemOpen
  \bibfield  {author} {\bibinfo {author} {\bibfnamefont {G.}~\bibnamefont
  {Sordi}}, \bibinfo {author} {\bibfnamefont {K.}~\bibnamefont {Haule}},\ and\
  \bibinfo {author} {\bibfnamefont {A.-M.~S.}\ \bibnamefont {Tremblay}},\
  }\bibfield  {title} {\bibinfo {title} {{Mott physics and first-order
  transition between two metals in the normal-state phase diagram of the
  two-dimensional Hubbard model}},\ }\href
  {https://doi.org/10.1103/PhysRevB.84.075161} {\bibfield  {journal} {\bibinfo
  {journal} {Phys. Rev. B}\ }\textbf {\bibinfo {volume} {84}},\ \bibinfo
  {pages} {075161} (\bibinfo {year} {2011})}\BibitemShut {NoStop}%
\bibitem [{\citenamefont {Sordi}\ \emph
  {et~al.}(2012{\natexlab{b}})\citenamefont {Sordi}, \citenamefont {S\'emon},
  \citenamefont {Haule},\ and\ \citenamefont {Tremblay}}]{ssht}%
  \BibitemOpen
  \bibfield  {author} {\bibinfo {author} {\bibfnamefont {G.}~\bibnamefont
  {Sordi}}, \bibinfo {author} {\bibfnamefont {P.}~\bibnamefont {S\'emon}},
  \bibinfo {author} {\bibfnamefont {K.}~\bibnamefont {Haule}},\ and\ \bibinfo
  {author} {\bibfnamefont {A.-M.~S.}\ \bibnamefont {Tremblay}},\ }\bibfield
  {title} {\bibinfo {title} {{Pseudogap temperature as a Widom line in doped
  Mott insulators}},\ }\href {https://doi.org/doi:10.1038/srep00547} {\bibfield
   {journal} {\bibinfo  {journal} {Sci. Rep.}\ }\textbf {\bibinfo {volume}
  {2}},\ \bibinfo {pages} {547} (\bibinfo {year}
  {2012}{\natexlab{b}})}\BibitemShut {NoStop}%
\end{thebibliography}%

\onecolumngrid
\appendix*

\section{Appendix}
\begin{figure*}[h!]
\centering{
\includegraphics[width=0.94\linewidth]{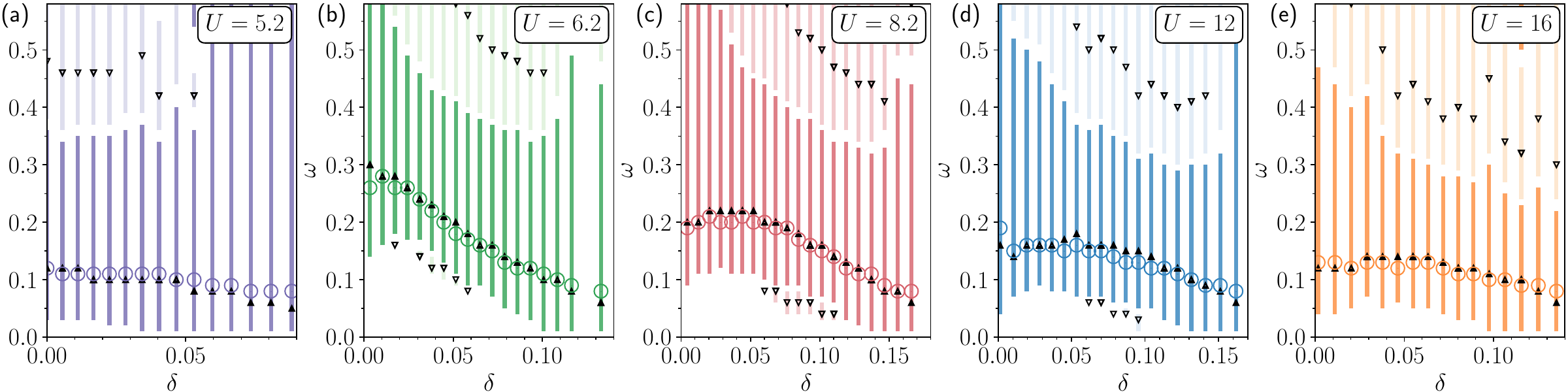}
}
\caption{Low-frequency zoom of Fig.~\ref{fig:freq-pairing}. For each value of $\delta$ and $U$, the frequency position of the low-frequency maximum of $A_{\rm an}(\omega)$ (filled up triangles) tracks the size of the superconducting gap $\Delta_{\rm sc}$ (open circles), suggesting that $\Delta_{\rm sc}$ is the energy scale where pairing is maximum~\cite{reymbautPRB2016}.
}
\label{fig:lowfreq-pairing}
\end{figure*}

Figure~\ref{fig:lowfreq-pairing} shows a zoom of Fig.~\ref{fig:freq-pairing} on the low frequencies. 
%



\clearpage

\begin{center}

{\bf Supplemental Material:} \\
\vspace{0.05cm} 

{\bf Dynamics of superconducting pairs in the two-dimensional Hubbard model} \\
\vspace{0.05cm}

{G. Sordi, E. M. O'Callaghan, C. Walsh, M. Charlebois, P. S\'emon, and A.-M. S. Tremblay}

\end{center}

\setcounter{figure}{0}
\renewcommand{\thefigure}{S\arabic{figure}}

\setcounter{subsection}{0}
\renewcommand{\thesubsection}{\arabic{subsection}}
\setcounter{secnumdepth}{3}

Here we describe how to calculate the anomalous spectral function $A_{\rm an}(\omega)$ by performing a single analytical continuation (Sec.~\ref{SMsec:Aan}) and the consistency checks for $A_{\rm an}(\omega)$ (Sec.~\ref{SMsec:consistencychecks}). Furthermore, Sec.~\ref{SMsec:spectra} shows extended data for the density of states in the superconducting state $A_{\rm nor}(\omega)$, the anomalous spectral function $A_{\rm an}(\omega)$, the cumulative spectral weight of the order parameter $I_{F}(\omega)$, and $A_{\rm an}(t)$ in the time domain.

\subsection{Obtaining the anomalous spectral function $A_{\rm an}(\omega)$}
\label{SMsec:Aan}

To compute the anomalous spectral function $A_{\rm an}(\omega)$ we start with the MaxEntAux method of Ref.~\cite{Alexis_PRB2015}. For the specific case of $d$-wave superconductivity with inversion symmetry and with time-reversal symmetry studied with CDMFT on a $2\times 2$ plaquette, this method involves defining the auxiliary Green’s function 
\begin{align}
G_{{\rm aux} \, {\bf K}} (\tau) & = - \langle T_{\tau} a_{{\bf K}}(\tau) a^\dagger_{{\bf K}}(0) \rangle , 
\end{align}
with $a_{{\bf K}} = c_{{\bf K}\uparrow} + c^\dagger_{-{\bf K}\downarrow}$ and ${\bf K}=(\pi,0)$. Introducing the Matsubara frequencies, one obtains~\cite{Alexis_PRB2015, reymbautPRB2016}
\begin{align}
G_{{\rm aux} \, {\bf K}} (i\omega_n) & = G_{{\bf K}\uparrow}(i\omega_n) - G_{{\bf K}\downarrow}(-i\omega_n) +2F_{{\bf K}}(i\omega_n) , 
\end{align}
where $G_{{\bf K}\sigma}(i\omega_n) = -\int_0^\beta d\tau e^{i\omega_n \tau} \langle T_\tau c_{{\bf K}\sigma}(\tau) c_{{\bf K}\sigma}^\dagger(0)\rangle$ is the Nambu diagonal (i.e. normal) Green’s function and $F_{{\bf K}}(i\omega_n) = -\int_0^\beta d\tau e^{i\omega_n \tau} \langle T_\tau c_{{\bf K}\uparrow}(\tau) c_{-{\bf K}\downarrow}(0)\rangle$ is the Nambu off-diagonal (i.e. anomalous) Green’s function. Using the maximum entropy software of Ref.~\cite{DominicMEM}, one can analytically continue $G_{{\bf K}\sigma}(i\omega_n)$ and $G_{{\rm aux} \, {\bf K}}(i\omega_n)$ to obtain $A_{{\bf K}\sigma}(\omega)$ and $A_{{\rm aux}\,{\bf K}}(\omega)$, and hence the anomalous spectral function
\begin{align}
\tilde{A}_{\rm an}(\omega) & = \frac{1}{2} \Big( A_{{\rm aux}\,{\bf K}}(\omega) -\big( A_{{\bf K}}(\omega) +A_{{\bf K}}(-\omega) \big) \Big) ,
\label{eqSM:Aan}
\end{align}
where $A_{{\bf K}}(\omega) = A_{{\bf K}\uparrow}(\omega) = A_{{\bf K}\downarrow}(\omega)$ and where $\tilde{A}_{\rm an}(\omega)  = \tilde{A}_{{\rm an} \,{\bf K}}(\omega) $ to simplify the notation. 

In our case (i.e., $d$-wave superconductivity with inversion symmetry and with time-reversal symmetry),  the physical $A_{\rm an}(\omega)$ can be shown~\cite{Alexis_PRB2015} to be real and odd in frequency. Numerically, the right hand side of Eq.~\ref{eqSM:Aan} is never perfectly odd in frequency, owing to the limitations of the analytical continuation. This is why we use the notation $\tilde{A}_{\rm an}(\omega)$.  
To overcome this problem, a simple method is to force the right hand side of Eq.~\ref{eqSM:Aan} to be odd in frequency by simply antisymmetrising it ($\tilde{A}_{\rm an, odd}(\omega) = \frac{1}{2}(\tilde{A}_{\rm an}(\omega)-\tilde{A}_{\rm an}(-\omega))$). 
Another equivalent method to obtain that $A_{\rm an} (\omega)$ is odd in frequency is to exploit the symmetries of the functions in Eq.~\ref{eqSM:Aan}. Indeed, first note that in Eq.~\ref{eqSM:Aan}, $A_{{\rm aux}\,{\bf K}}(\omega)$ is in general not even nor odd in frequency, whereas $(A_{{\bf K}}(\omega) +A_{{\bf K}}(-\omega))$ is even in frequency. Since any real function can be written as the sum of an even function and an odd function, we can write Eq.~\ref{eqSM:Aan}  as 
\begin{align}
\tilde{A}_{\rm an} (\omega) & = \tilde{A}_{\rm an, odd} (\omega) +\tilde{A}_{\rm an, even} (\omega), 
\label{eqSM:Aan2}
\end{align}
where $\tilde{A}_{\rm an, odd} (\omega)= \tfrac12 A_{\rm aux\,{\bf K}, odd} (\omega)$ and $\tilde{A}_{\rm an, even} (\omega) = \tfrac12 A_{\rm aux \,{\bf K}, even} (\omega) - \tfrac12 ( A_{{\bf K}}(\omega) +A_{{\bf K}}(-\omega) )$. 

Antisymmetrizing $\tilde{A}_{\rm an}(\omega)$ or $A_{\rm aux\,{\bf K}}(\omega)$ is equivalent. However, numerically, the latter is advantageous since it allows us to perform a {\it single} analytical continuation (of $G_{{\rm aux} \, {\bf K}}(i\omega_n)$) rather than two analytical continuations (of $G_{{\bf K}\sigma}(i\omega_n)$ and $G_{{\rm aux} \, {\bf K}}(i\omega_n)$). 
In practice, first we analytically continue $G_{{\rm aux} \, {\bf K}}(i\omega_n)$ to obtain $A_{{\rm aux}\,{\bf K}}(\omega)$. Second, we antisymmetrize $A_{\rm aux\,{\bf K}} (\omega)$, i.e. $A_{\rm aux\,{\bf K}, odd}(\omega) = \frac{1}{2}(A_{\rm aux\,{\bf K}}(\omega)-A_{\rm aux\,{\bf K}}(-\omega))$, which is exactly $2\tilde{A}_{\rm an, odd} (\omega)$.

\begin{figure*}[t!]
\centering{
\includegraphics[width=0.999\linewidth]{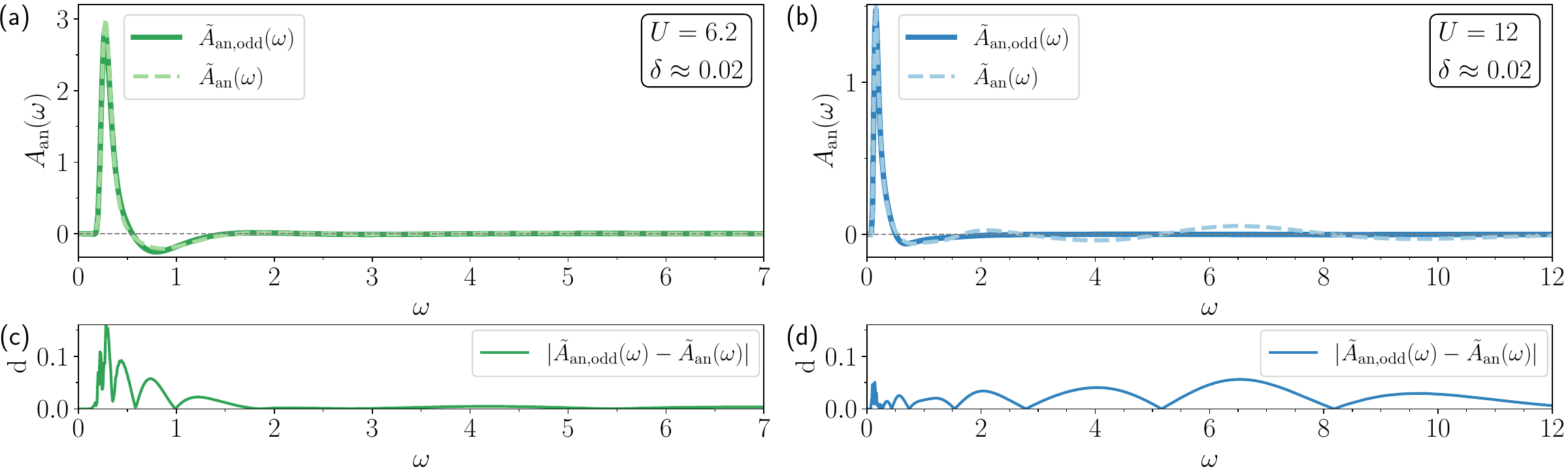}
}
\caption{Panels (a) and (b): numerical approximations for the physical $A_{\rm an}(\omega)$: $\tilde{A}_{\rm an, odd}(\omega)$, obtained with antisymmetrization of  $\tfrac12 A_{\rm aux\,{\bf K}} (\omega)$ (and a single analytical continuation), and $\tilde{A}_{\rm an}(\omega)$ obtained without antisymmetrization (and two analytical continuations), for $U=6.2, \delta\approx 0.02$ and $U=12, \delta\approx 0.02$. Panels (c) and (d): absolute distance $d = |\tilde{A}_{\rm an,odd}(\omega) - \tilde{A}_{\rm an}(\omega)|$. In the main text we use $A_{\rm an}(\omega) = \tilde{A}_{\rm an, odd}(\omega)= \tfrac12 A_{\rm aux\,{\bf K}, odd} (\omega)$. }
\label{figSM:Aan}
\end{figure*}

Next, in Figure~\ref{figSM:Aan} we compare $\tilde{A}_{\rm an,odd}(\omega)$ and $\tilde{A}_{\rm an}(\omega)$ for two representative cases: $U=6.2, \delta\approx 0.02$ and $U=12, \delta\approx 0.02$. Panels (c) and (d) show the absolute distance between $\tilde{A}_{\rm an,odd}(\omega)$ and $\tilde{A}_{\rm an}(\omega)$. 
Note that no antisymmetrization of $\tilde{A}_{\rm an}(\omega)$ gives rise to some fluctuations at high frequency, which are more pronounced for $U=12$. This is because Eq.~\ref{eqSM:Aan} is based on the subtraction of two analytically continued spectra which have structures at frequency of order $U$ related to the development of the Hubbard bands. Since the analytical continuation degrades with higher frequencies, this subtraction can give rise to fluctuations at high frequency which becomes more important with increasing $U$. Antisymmetrization cure this shortcoming at higher frequencies.  
In the main text and for the rest of the supplemental material, we use $\tilde{A}_{\rm an,odd}(\omega)$ as a numerical proxy for the physical $A_{\rm an}(\omega)$. 

\subsection{Consistency checks for $A_{\rm an}(\omega)$}
\label{SMsec:consistencychecks}

It is important to perform some consistency checks for the analytically continued anomalous spectral function $A_{\rm an}(\omega)$. A stringent consistency check is provided by the fact that in the limit $\omega \rightarrow \infty$, the cumulative spectral weight of the order parameter 
\begin{align}
I_F(\omega) & =\int_{-\omega}^{\omega} \frac{d\omega'}{2\pi} A_{\rm an} (\omega') f({-\omega'})
\label{SMeq:IF}
\end{align}
converges to the superconducting order parameter $|\Phi|$. 

\begin{figure*}[ht!]
\centering{
\includegraphics[width=0.999\linewidth]{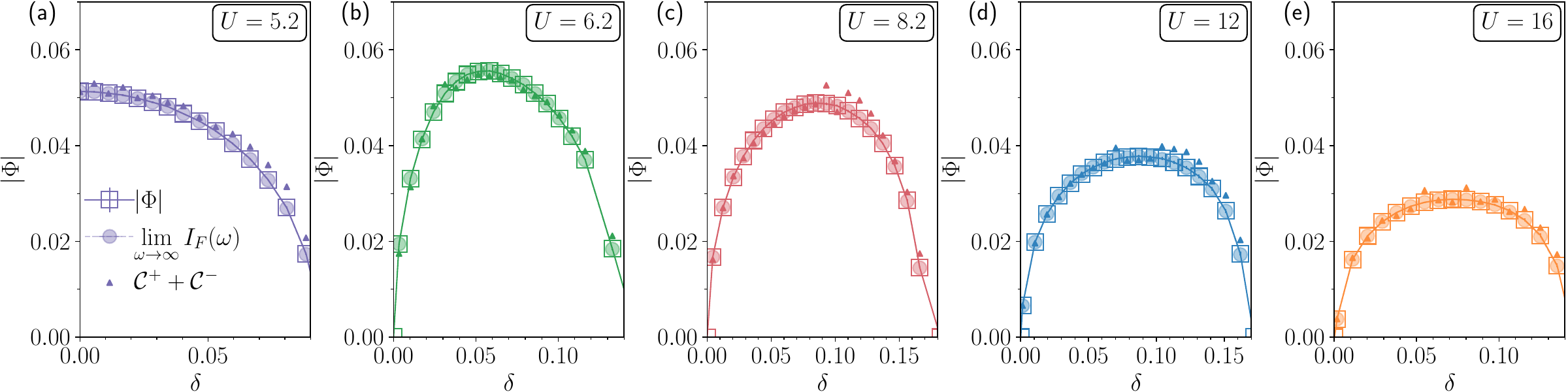}
}
\caption{Superconducting order parameter $|\Phi|$ at $T=1/50$ as a function of doping $\delta$ and for different values of $U$, computed in two independent ways: (i) directly within the impurity solver, from $\Phi=\langle F_{{\bf K}=(\pi,0)}(\tau=0^{+}) \rangle$ (open squares) and (ii) from $\lim \limits_{\omega \rightarrow \infty} I_F(\omega)$ (filled circles). The latter has been estimated as the average over the frequency range $\omega \in [30, 40]$ of $I_F(\omega)$. Filled triangles denote the sum of the largest positive and negative area of $A_{\rm an}(\omega)$, $\mathcal{C}^+ +\mathcal{C}^-$. Overall, they add up to $|\Phi|$. The slight deviations from $|\Phi|$ are likely due to the limitations of the analytical continuation and signal small pairing and depairing contributions at larger frequency scales. 
}
\label{fig:SCop}
\end{figure*}

Figure~\ref{fig:SCop} shows $|\Phi|$ as a function of doping and for different values of $U$, computed in two independent ways: (i) directly within the impurity solver, from $\Phi=\langle F_{{\bf K}=(\pi,0)}(\tau=0^{+}) \rangle$ (open squares), i.e. without analytical continuation and (ii) from $\lim \limits_{\omega \rightarrow \infty} I_F(\omega)$ (filled circles). The former does not rely on analytical continuation, whereas the latter depends on it. The excellent agreement between the two protocols for all values of $U$ and $\delta$ indicates that the main features of $A_{\rm an}(\omega)$ are correctly captured by the analytical continuation. 

Another consistency check is provided by the good agreement between our analytically continued results for $I_F(\omega)$ at $U=8.2$ (see Fig.~\ref{fig:SMspectra-u82}) and the zero temperature data at $U=8$ (and nonzero next-nearest neighbor hopping) of Ref.~\cite{Kyung:2009} which are obtained directly on the real frequencies without analytical continuation. On the other hand, our data appear incompatible with the results of Ref.~\cite{maierPRL2008} based on DCA and a non-crossing approximation as impurity solver, which, contrary to our findings, show an increased importance of the high-frequencies pairing forming process with increasing $U$ in the range $8<U<12$. The results of Ref.~\cite{Gull:PRB2014} at $U=6$ based on DCA and a continuous-time auxiliary field impurity solver are compatible with the low-frequencies pairing forming process found in our work, but, contrary to our findings, they do not show pair-breaking processes.

\subsection{Extended data for $A_{\rm nor}(\omega)$, $A_{\rm an}(\omega)$, $I_{F}(\omega)$, and $A_{\rm an}(t)$}
\label{SMsec:spectra}
\begin{figure*}
\centering{
\includegraphics[width=0.99\linewidth]{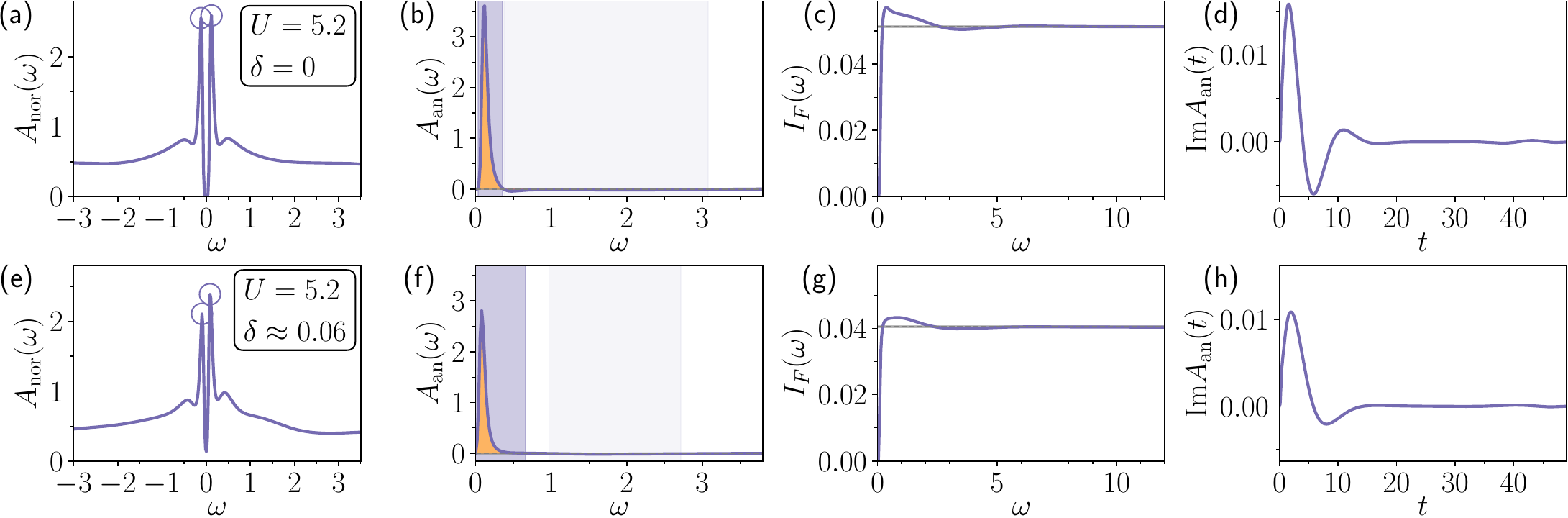}
}
\caption{(a, e) Density of states in the superconducting state $A_{\rm nor}(\omega)$. The open circles mark the superconducting coherence peaks. (b, f) Anomalous spectral function $A_{\rm an}(\omega)$. The frequency regions where $A_{\rm an}(\omega)$ is positive (negative) are marked by dark (pale) colored vertical bands and the corresponding positive (negative) areas are colored in yellow (red). (c, g) Cumulative spectral weight of the order parameter $I_F(\omega)$. Horizontal line indicates the superconducting order parameter $|\Phi|$ calculated independently and directly within the impurity solver. (d, h) Imaginary part of the anomalous spectral function $A_{\rm an}(t)$ in the time domain. Data are for $U=5.2$, $T=1/50$ and $\delta=0$ (top row) and $\delta\approx 0.06$ (bottom row).
}
\label{fig:SMspectra-u52}
\end{figure*}

\begin{figure*}
\centering{
\includegraphics[width=0.99\linewidth]{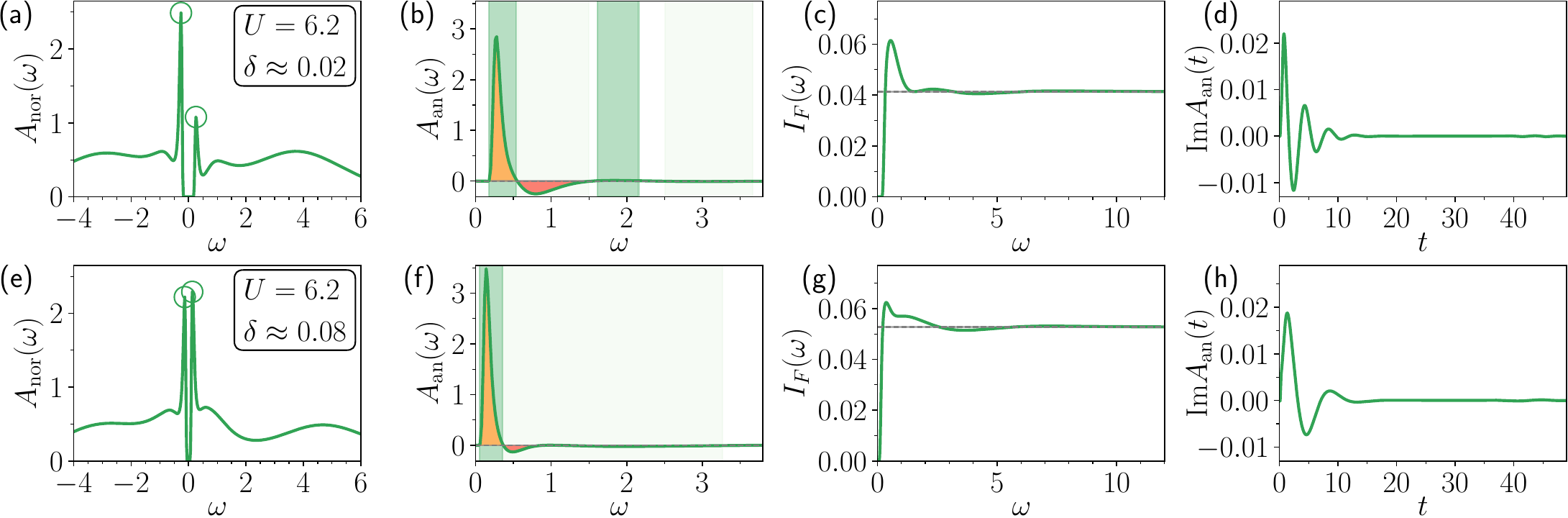}
}
\caption{Same as Fig.~\ref{fig:SMspectra-u52}, but for $U=6.2$, $T=1/50$ and $\delta \approx 0.02$ (top row) and $\delta\approx 0.08$ (bottom row).
}
\label{fig:SMspectra-u62}
\end{figure*}

\begin{figure*}
\centering{
\includegraphics[width=0.99\linewidth]{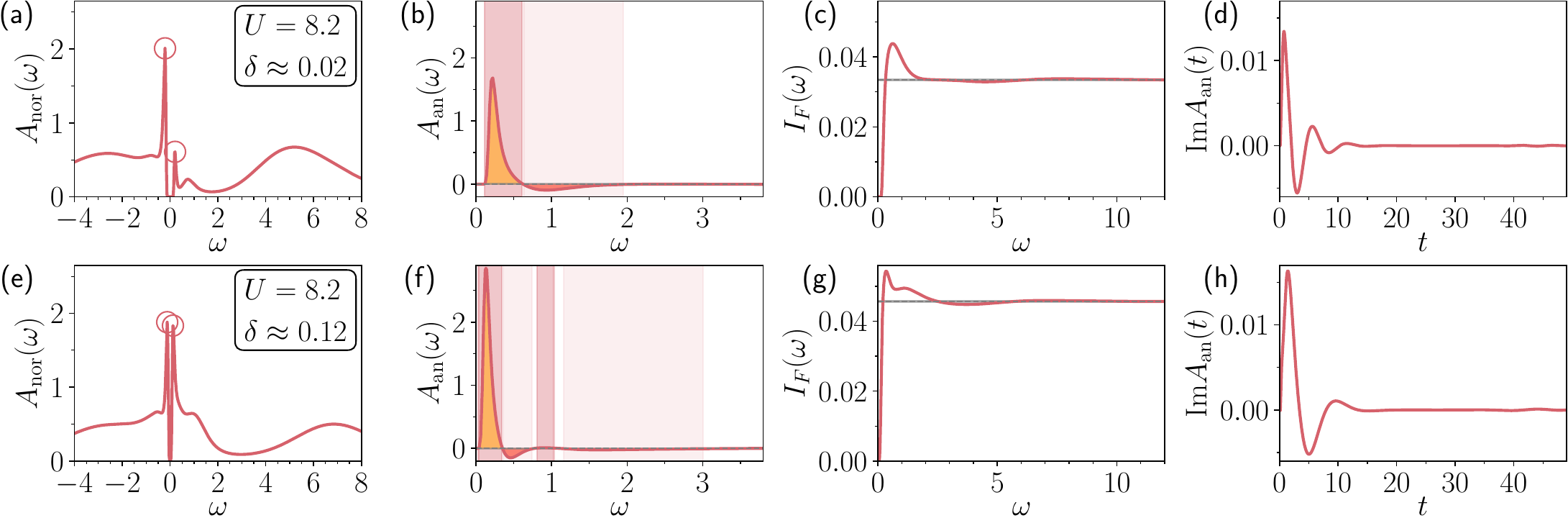}
}
\caption{Same as Fig.~\ref{fig:SMspectra-u52}, but for $U=8.2$, $T=1/50$ and $\delta \approx 0.02$ (top row) and $\delta\approx 0.12$ (bottom row).
}
\label{fig:SMspectra-u82}
\end{figure*}

\begin{figure*}
\centering{
\includegraphics[width=0.99\linewidth]{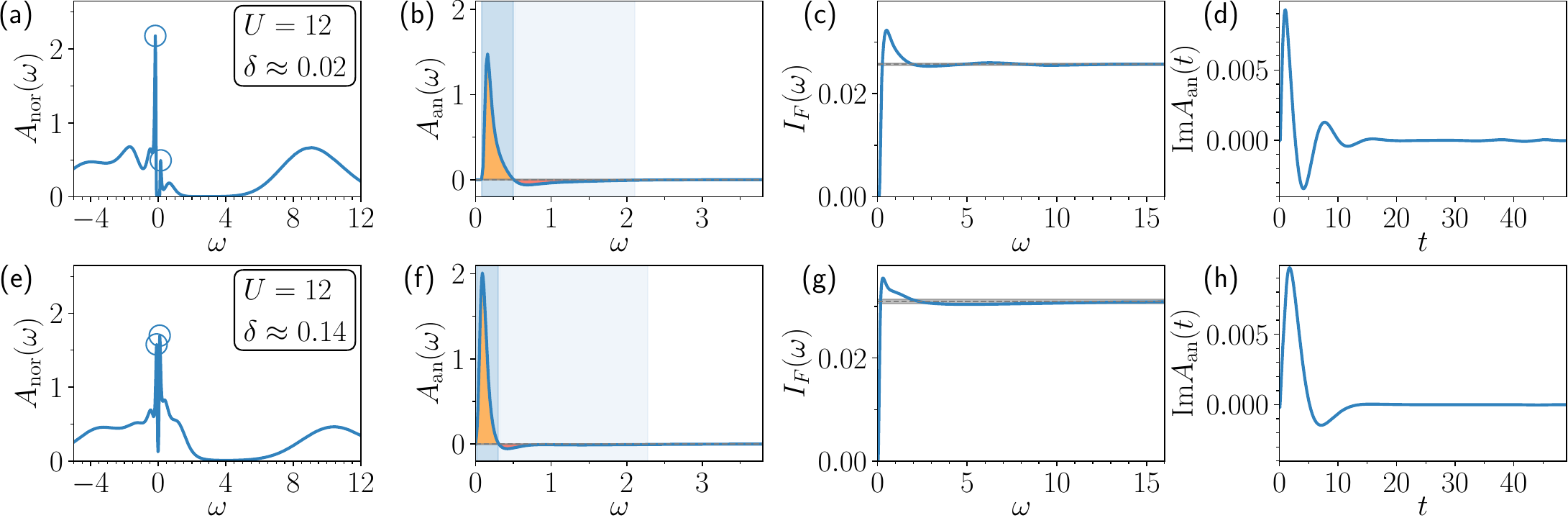}
}
\caption{Same as Fig.~\ref{fig:SMspectra-u52}, but for $U=12$, $T=1/50$ and $\delta \approx 0.02$ (top row) and $\delta\approx 0.14$ (bottom row).
}
\label{fig:SMspectra-u12}
\end{figure*}

\begin{figure*}
\centering{
\includegraphics[width=0.99\linewidth]{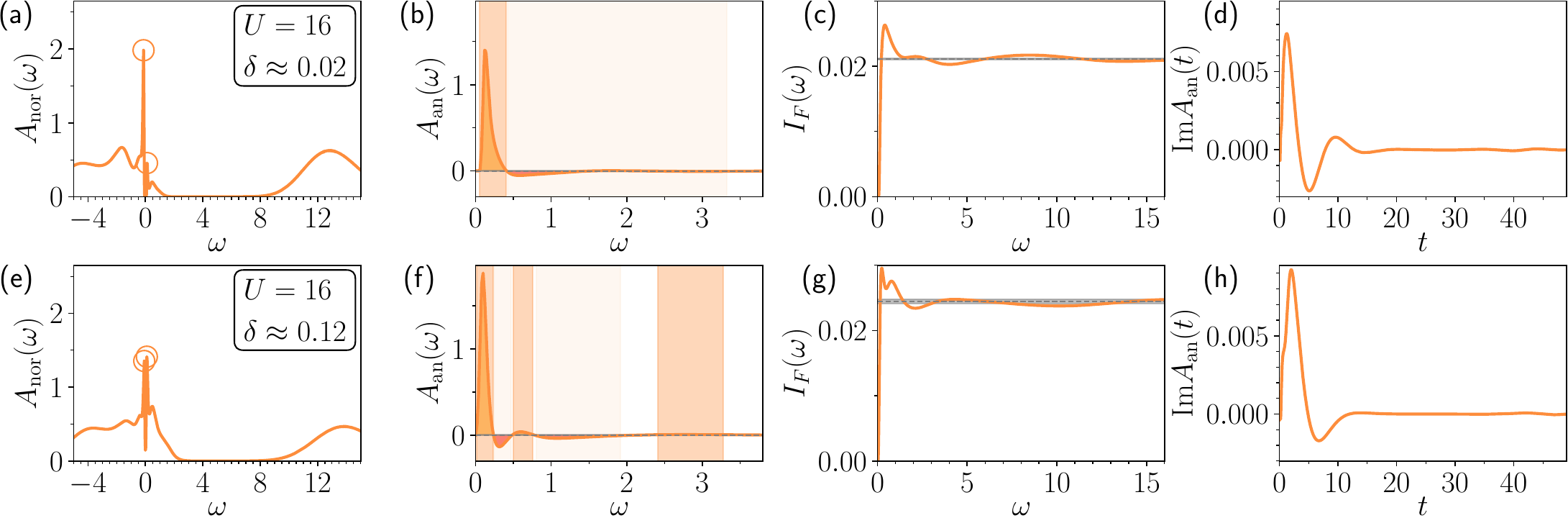}
}
\caption{Same as Fig.~\ref{fig:SMspectra-u52}, but for $U=16$, $T=1/50$ and $\delta \approx 0.02$ (top row) and $\delta\approx 0.12$ (bottom row).
}
\label{fig:SMspectra-u16}
\end{figure*}

With the same notation of Fig.~1 of the main text, we show in Figs.~\ref{fig:SMspectra-u52} - \ref{fig:SMspectra-u16} the density of states in the superconducting state $A_{\rm nor}(\omega)$, the anomalous spectral function $A_{\rm an}(\omega)$, and the cumulative spectral weight of the order parameter $I_{F}(\omega)$, for different values of $U$ and doping levels $\delta$. The color code for each value of $U$ is the same as in the main text. 

Specifically, the frequency regions where $A_{\rm an}(\omega)$ is positive (negative) are those that contribute to pairing (depairing): they are shaded with dark (pale) colored vertical bands and are examined in Fig.~3 of main text. To calculate these frequency regions, we introduce a small threshold ($|A_{\rm an}(\omega)| <0.005$, see horizontal grey interval in panels (b),(f) of  Figs.~\ref{fig:SMspectra-u52} - \ref{fig:SMspectra-u16}) with the aim of filtering out the noise in the analytically continued $A_{\rm an}(\omega)$ spectra. The blank bands in panels (b),(f) of  Figs.~\ref{fig:SMspectra-u52} - \ref{fig:SMspectra-u16} indicate the frequency regions where $A_{\rm an}(\omega)$ is negligible. For all valued of $U$ and $\delta$, $A_{\rm an}(\omega)$ is negligible at high frequencies (i.e. $\omega \gtrsim 6$). At low frequencies, this threshold may give rise to a blank band between dark and pale bands (dark and pale bars in Fig.~3 of main text). 

The areas under (above) $A_{\rm an}(\omega)$ and above (under) the small threshold of $\pm 0.005$ are colored in yellow (red) in panels (b),(f) of Figs.~\ref{fig:SMspectra-u52} - \ref{fig:SMspectra-u16}. With the composite trapezoidal method, we compute the largest positive and negative area for each value of $U$ and $\delta$ (denoted $\mathcal{C}^+$ and $\mathcal{C}^-$ in main text) and we analyse their behavior versus $\delta$ and $U$ in Fig.~4 of main text. By inspection, these largest areas occur at low frequencies, for all values of $U$ and $\delta$. Assuming this is true, then the sum $\mathcal{C}^+ +\mathcal{C}^-$ should be approximately equal to the superconducting order parameter $|\Phi|$. This is because at the low temperature $T=1/50$ studied in our work, $I_F(\omega)$ defined in Eq.~\ref{SMeq:IF} is approximately the integral of $A_{\rm an}(\omega)$ over the positive frequencies, i.e. $I_F(\omega) \approx \int_{0}^{\omega} \frac{d\omega'}{2\pi} A_{\rm an} (\omega')$. This is indeed the case, as shown in Fig.~\ref{fig:SCop} where $\mathcal{C}^+ +\mathcal{C}^-$ (filled triangles) overall add up to $|\Phi|$.

In addition, Figs.~\ref{fig:SMspectra-u52} - \ref{fig:SMspectra-u16} show the anomalous spectral function $A_{\rm an}(t)$ in the time domain, computed with the fast-Fourier transform of $A_{\rm an}(\omega)$. Note that since $A_{\rm an}(\omega)$ is real and odd in frequency, then $A_{\rm an}(t)$ is purely imaginary. The  key features of $\textrm{Im} A_{\rm an}(t)$ in the time domain can be qualitatively understood by modeling the positive frequency part of $A_{\rm an}(\omega)$ as the sum of two signed delta functions centered at frequency $\omega_1$ and $\omega_2$, each of which is convoluted by a Gaussian function. Linearity of the Fourier transform implies that $\textrm{Im} A_{\rm an}(t)$ is the sum of two sinus functions of frequency $\omega_1$ and $\omega_2$, each of which multiplied by a Gaussian centered at $t=0$, resulting in a decaying sinusoidal function versus time. Furthermore, reducing the doping level shifts the low-frequency peak in $A_{\rm an}(\omega)$ to higher frequencies, resulting in faster oscillations of $\textrm{Im} A_{\rm an}(t)$ in the time domain.


\end{document}